\documentstyle[aps,eqsecnum,preprint]{revtex}

\begin{document}
\newcommand\1{$\spadesuit$}
\newcommand\2{$\clubsuit$}
\tighten
\preprint{ETH-TH/97-7}
\draft
\title{The influence of cosmological transitions on the evolution 
of density perturbations}
\author{J\'er\^ome Martin\thanks{e-mail: martin@edelweiss.obspm.fr}}
\address{DARC, Observatoire de Paris, \\ 
UPR 176 CNRS, 92195 Meudon Cedex, France.}
\author{Dominik J. Schwarz
\thanks{e-mail: dschwarz@th.physik.uni-frankfurt.de}
\thanks{Address since October 1, 1997: Institut f\"ur Theoretische Physik,
Robert-Mayer-Stra\ss e 8 -- 10, Postfach 11 19 32, D-60054 Frankfurt am Main}}
\address{Institut f\"ur Theoretische Physik, 
ETH-H\"onggerberg, CH-8093 Z\"urich}

\date{November 26, 1997}
\maketitle

\begin{abstract}
We study the influence of the reheating and equality transitions 
on superhorizon density perturbations and gravitational waves. 
Recent criticisms of the `standard result' for large-scale 
perturbations in inflationary cosmology are rectified. 
The claim that the `conservation law' for the amplitude of superhorizon 
modes was empty is shown to be wrong. For sharp transitions, 
i.e.~the pressure jumps, we rederive the Deruelle-Mukhanov junction 
conditions. For a smooth transition we correct a result obtained
by Grishchuk recently. We show that the junction conditions are not crucial,
because the pressure is continuous during the reheating transition. The 
problem occurred, because the perturbed metric was not evolved correctly
through the smooth reheating transition. Finally, we derive the `standard
result' within Grishchuk's smooth (reheating) transition.
\end{abstract}

\pacs{PACS numbers: 98.80.Cq, 98.70.Vc}

\section{Introduction}

The discovery of anisotropies in the Cosmic Microwave Background 
Radiation (CMBR) \cite{Smooth} was a major advance for Cosmology. 
Presently, a series of experiments is being conducted to 
measure these anisotropies on smaller angular scales. It is likely that 
the CMBR anisotropies origin from primordial fluctuations of the metric 
and matter fields which have been amplified later on. Therefore, the outcomes 
of the ongoing and planned observations may have important consequences 
on the theoretical models explaining the origin and the evolution of 
cosmological perturbations, and more generally, on our ideas about 
the early Universe.
\par
Among these ideas is the theory of perturbations generated 
quantum-mechanically during an inflationary stage. In the present work
our attention shall be restricted to inflationary models with one 
scalar field. In the subsequent evolution of the Universe 
matter shall be described
by a perfect fluid. With the help of various models definite predictions 
for the amplitudes and spectra of the different types of 
perturbations (density perturbations, rotational perturbations and 
gravitational waves) can be made. In particular, allowing perfect
fluids only, one can show that rotational perturbations have to decay.
Moreover, single scalar field inflation cannot seed rotational perturbations. 
\par
The power spectrum of density perturbations in an 
inflationary Universe was first calculated by Mukhanov and 
Chibisov \cite{MuChi}. This first computation was confirmed in 
Refs. \cite{Haw,Staro,GuPi,BST}. Essentially, the `standard result' lies 
in the following: the closer the spectrum is to the flat 
(Harrisson-Zeldovich) spectrum, the larger the amplitude of density 
perturbations is in comparison with the amplitude of gravitational 
waves \cite{gw,Smoot}. On the other hand, if the spectrum is 
tilted away from scale-invariance, the contribution due to gravitational 
waves can become important.
(For some specific examples of this behavior see, e.g.~\cite{T}.)
\par
This result was recently challenged by Grishchuk in Ref. \cite{G1} who 
found a similar amplitude for density perturbations and 
gravitational waves. However, his 
calculations were criticized by Deruelle and Mukhanov \cite{DM}. They 
argued that Grishchuk did not properly consider the joining 
conditions at the transitions `inflation-radiation' and `radiation-matter'. 
Then, Grishchuk published a comment \cite{G2} in which he stated that the 
equation expressing the `standard result' is mathematically wrong. 
Short after, this claim was contested by Caldwell \cite{C} who 
expressed his agreement with Deruelle and Mukhanov. Finally, in 
the appendix of Ref. \cite{G3}, Grishchuk re-stated his criticisms 
of the `standard result' in more details.
\par
The aim of this paper is to clear up this controversy and to study 
the influence of cosmological transitions undergone by the Universe  
during its history on superhorizon cosmological perturbations. 
\par
This article is organized as follows: the second section is 
devoted to density perturbations and gravitational waves in the 
gauge-invariant formalism. This section can be 
skipped by specialists. The third section deals with the 
synchronous-gauge formalism. It was claimed by Grishchuk that the synchronous 
gauge and gauge-invariant results differ by an arbitrary constant of 
integration. Thus, a systematic comparison of the two formalisms is 
made and we show that they are equivalent. In the 
fourth section, we study the so-called `conservation law' for 
superhorizon modes. We show that the `conserved quantity' $\zeta$ often 
used in the literature is actually not empty as it was 
claimed in Refs. \cite{G2}. In the fifth section, we turn to the question 
of the matching conditions. In particular, we re-derive the 
Deruelle-Mukhanov junction conditions with a different method. The 
sixth section is devoted to the study of the smooth transition of 
Ref.~\cite{G1}. We will argue that the joining conditions are not the essential
point, but that the evolution of the metric perturbations through the 
smooth reheating transition was not done correctly. We correct the result,
which is now the same as for a sharp transition.
In the last section, we briefly present our conclusions, which 
confirm the `standard result'. Finally, an appendix reviews how the 
initial conditions are fixed when the perturbations are quantum-mechanically 
generated. For the readers convenience we summarize 
the notation of the gauge-invariant formulation and the synchronous-gauge 
formulation in two tables at the end of the paper.

\section{Gauge-invariant scalar and tensor perturbations}

We will use the notation of \cite{MFB} except the signature of the metric 
and the harmonic decomposition of the gauge-invariant potentials. 
The line element for the Friedmann-Lemaitre-Robertson-Walker (FLRW) background
plus scalar perturbations reads ($c=1$)
\begin{equation}
\label{ds2}
{\rm d}s^2 = a(\eta )^2\{- (1+2\phi){\rm d}\eta ^2 + 2B_{|i}{\rm d}x^i
{\rm d}\eta + [(1-2\psi)\gamma _{ij}+2E_{|i|j}]{\rm d}x^i{\rm d}x^j\} \ . 
\end{equation}
The conformal time $\eta$ is related to the cosmic time $t$ by
${\rm d}t = a(\eta){\rm d}\eta$. A dot denotes a derivative with respect to  
$t$, whereas a prime stands for a derivative with respect to $\eta$. The 
tensor $\gamma _{ij}$ is the metric of the three-dimensional space-like 
hypersurfaces and the derivative ${}_|$ in (\ref{ds2}) is covariant with 
respect to $\gamma_{ij}$. The curvature of these sections 
is given by $K$ and takes the values $0$, $\pm 1$.  

One can generate fictitious perturbations by 
performing an infinitesimal change of coordinates which preserves the 
scalar form of equation (\ref{ds2}) (let us note that 
this is not, as written in \cite{C} on page 2439, 
a ``general'' infinitesimal transformation of coordinates):
\begin{equation}
\label{2-6}
\bar{\eta }=\eta +\xi ^0(\eta ,x^k), \qquad  
\bar{x}^i=x^i+\gamma ^{ij}\xi _{|j}(\eta ,x^k) \ .
\end{equation}
Two independent gauge-invariant scalar metric potentials 
may be constructed from the metric. Following Ref.~\cite{MFB} we take them
to be:
\begin{equation}
\label{gimp}
\Phi Q \equiv \phi +\frac{1}{a}[(B-E')a]', \qquad
\Psi Q \equiv \psi -\frac{a'}{a}(B-E').
\end{equation}
In these expressions, we have chosen to extract the scalar harmonic $Q(x^i)$ 
of the gauge-invariant variables from the very beginning. The function $Q(x^i)$ satisfies the Helmholtz equation 
\begin{equation}
\bigtriangleup Q = - k^2 Q \ ,
\end{equation} 
where $\bigtriangleup Q \equiv \gamma^{ij} Q_{|i|j}$ and $k$ is the 
comoving wave number. For convenience the quantity
${\cal H} \equiv a'/a$ is defined, which is related to the Hubble
parameter $H \equiv \dot{a}/a = {\cal H}/a$. 
The perturbed Einstein equations can be expressed in terms of 
$\Phi $ and $\Psi$ alone (see below). Working in the gauge-invariant 
formulation of Bardeen \cite{B1} is equivalent to the longitudinal
gauge ($B=E=0$), which fixes the constant time hypersurfaces to be the
hypersurfaces with vanishing shear.

In most inflationary models matter is described by a scalar field  
$\varphi = \varphi_0 (\eta ) + \delta\varphi (\eta,x^i)$. The 
background energy density and 
pressure are
\begin{equation}
\label{2-2}
\rho_0=\frac{(\varphi _0')^2}{2a^2}+V(\varphi _0), \qquad 
p_0=\frac{(\varphi _0')^2}{2a^2}-V(\varphi _0) \ ,
\end{equation}
where $V(\varphi)$ is the potential of the scalar field.
If $\varphi_0' \neq 0$,
the covariant conservation of the energy-momentum tensor 
provides the Klein-Gordon equation
\begin{equation}
\label{2-3}
\varphi _0''+2{\cal H}\varphi _0'+a^2 V_{,\varphi}=0 \ .
\end{equation}
The gauge-invariant scalar field perturbation is 
$\delta\varphi^{\rm (gi)} Q \equiv \delta\varphi + \varphi'_0(B-E') $.
The linearized Einstein equations for a fixed mode $k$ may be written
in terms of gauge-invariant quantities only:
\begin{eqnarray}
\label{2-11}
& & -3{\cal H}({\cal H}\Phi +\Psi') - k^2 \Psi + 3K\Psi =
\frac{\kappa }{2}[-(\varphi _0')^2 \Phi + 
\varphi _0'(\delta\varphi^{\rm (gi)})'+
a^2 V_{,\varphi}\delta\varphi^{\rm (gi)}] \ , \\
\label{2-12}
& & {\cal H}\Phi +\Psi' = \frac{\kappa }{2} \varphi _0'
\delta\varphi^{\rm (gi)} \ , \\
\label{2-13}
& & \Phi - \Psi = 0 \ , \\
\label{2-14}
& & (2{\cal H}'+{\cal H}^2)\Phi +
{\cal H}\Phi '+\Psi ''+2{\cal H}\Psi' - K\Psi -
\frac{1}{3} k^2 (\Phi -\Psi) = \nonumber \\
& & \quad 
\frac{\kappa }{2}[-(\varphi _0')^2 \Phi +
\varphi _0'(\delta \varphi ^{(gi)})'-
a^2 V_{,\varphi}\delta \varphi ^{(gi)}] \ ,
\end{eqnarray}
where $\kappa \equiv 8\pi G$.
Everything can be expressed in terms of a single 
gauge-invariant quantity since equation (\ref{2-13}) tells us that 
$\Phi =\Psi$. 

If $\varphi _0'=0$ the background solution is 
the de Sitter spacetime ($\rho_0 = -p_0$). In that case, the solution 
of the system (\ref{2-11}) -- (\ref{2-14}) is $\Phi =0$: there 
are no scalar metric perturbations. This does not mean that 
the scalar field cannot fluctuate but that these fluctuations do 
not couple to gravity.

For $\varphi _0'\neq 0$ the linearized Einstein equations (\ref{2-11}) 
-- (\ref{2-14}) reduce with 
help of the Friedmann equation and the Klein-Gordon equation (\ref{2-3})
to the equation \cite{MFB}
\begin{equation}
\label{2-15}
\Phi'' + 2({\cal H}-\frac{\varphi _0''}{\varphi _0'})\Phi' +
[k^2 + 2({\cal H}'-{\cal H}\frac{\varphi _0''}
{\varphi _0'} - 2 K)]\Phi = 0 \ .
\end{equation}
For $K = 0$, the introduction of the new variables\footnote{The perturbation
$\sigma$ is related to $u$ of Ref. \cite{MFB} by $\sigma =
(\kappa/3)^{1/2} u$.} 
\begin{equation} 
\label{sigma} 
\sigma\equiv\frac23{a^2\theta\over{\cal H}}\Phi \ ,
\qquad \theta \equiv \frac1a \left(\rho_0\over \rho_0 + p_0\right)^{\frac12}
= \left({3\over\kappa}\right)^{\frac12} {{\cal H}\over a\varphi _0'} \ ,
\end{equation}
allows us to express equation (\ref{2-15}) in the form:
\begin{equation}
\label{2-16}
\sigma''+(k^2-\frac{\theta ''}{\theta })\sigma=0 \ .
\end{equation}

For modes $k^2 \ll \theta''/\theta$ the solution of Eq.~(\ref{2-16})
may be expanded in powers of $k^2$. At the leading orders we obtain 
\begin{eqnarray}
\label{sigmaexp}
\sigma &=& \tilde{A}_1(k) \theta\int{1\over \theta^2} \left(1-k^2
\int^\eta \theta^2 
\int^{\bar{\eta}} {1\over \theta^2}{\rm d} \tilde{\eta} {\rm d}\bar{\eta} +
{\cal O}(k^4)\right) {\rm d}\eta + \\
&& \tilde{A}_2(k) \theta\left(1 - k^2 \int{1\over\theta^2}\int^\eta 
\theta^2 {\rm d}\bar{\eta} {\rm d}\eta + {\cal O}(k^4)\right) \ . \nonumber
\end{eqnarray}
Since $\theta \to \infty$ for $a \to 0$ in general,
$\tilde{A}_1$ is the arbitrary constant 
in front of the regular (growing) mode and $\tilde{A}_2$ a constant 
associated with the singular (decaying) mode\footnote{
An example where both modes are singular as $a \to 0$ is provided by 
the model with scale factor behavior $a \propto |\eta|^{1+\beta}$, 
for $-2 <\beta < -1$ and $\eta<0$. However, this inflationary model 
violates the weak energy condition $\rho_0 + p_0 \geq 0$ and cannot be 
realized with a single real scalar field.}. 

For gravitational waves the line element reads:
\begin{equation}
\label{ds2t}
{\rm d}s^2 = a(\eta )^2\{-{\rm d}\eta^2 + 
[\gamma_{ij} + h_{ij}]{\rm d}x^i{\rm d}x^j\} \ , 
\end{equation}
the tensor $h_{ij}$ being symmetric, traceless and transverse. The
tensor sector is gauge-invariant. We write $h_{ij} = h_{\rm gw} Q_{ij}$,
where $Q_{ij}$ is a symmetric, transverse, and traceless spherical harmonic,
and $h_{\rm gw}$ is the amplitude of a gravitational wave.
The decay of the amplitude due to the expansion of the Universe is taken 
into account by defining $\mu _{\rm gw} \equiv a(\eta) h_{\rm gw}(\eta )$.
The equation of motion for $\mu_{\rm gw}$ reads: 
\begin{equation}
\label{eomgw}
\mu_{\rm gw}'' + (k^2 - \frac{a''}{a})\mu_{\rm gw} = 0 \ . 
\end{equation}
A physical interpretation of Eqs.~(\ref{2-16}) and (\ref{eomgw}) is 
parametric amplification of the perturbations while the Universe is 
expanding \cite{Gpara}. The scale factor plays the role of a `pump field' 
and the `interaction' between the background and the perturbations is 
described 
by the `potentials' $\theta''/\theta$ and $a''/a$. The potential of the
scalar perturbations involves not only 
the scale factor but also the derivatives of $a(\eta)$ up to $a^{(iv)}(\eta)$.

For simple models where the scale factor is given by 
$a(\eta )=l_0|\eta |^{1+\beta }$, the exact solution to Eq. (\ref{eomgw}) 
can be found. It reads:
\begin{equation}
\label{solgw}
\mu_{\rm gw} =(k\eta )^{1/2}[A_1^{\rm gw}(k) J_{\beta +1/2}(k\eta )
+A_2^{\rm gw}(k) J_{-(\beta +1/2)}(k\eta )],
\end{equation}
where $J_{\pm (\beta +1/2)}$ are Bessel functions.

For an arbitrary scale factor, as for the scalar metric perturbations, the 
solution for modes $k^2 \ll a''/a$ is given for gravitational waves 
by replacing $\theta$ with $a$ in Eq.~(\ref{sigmaexp}):
\begin{equation}
\label{approgw}
\mu_{\rm gw} = \bar{A}_1^{\rm gw}(k) a +\bar{A}_2^{\rm gw}(k)
a \int^{\eta }\frac{{\rm d}\bar{\eta}}{a^2} +{\cal O}(k^2).
\end{equation}
Now $\bar{A}_1^{\rm gw}$ corresponds to the regular and $\bar{A}_2^{\rm gw}$ 
to the singular mode. It should be noticed that the $A^{\rm gw}$'s differ 
from the $\bar{A}^{\rm gw}$'s, although they are connected.
At superhorizon scales the dominant mode is constant in time,
independent of the matter content of the Universe.

\section{Scalar perturbations in synchronous gauge}

Let us describe the scalar perturbations using the class of the 
synchronous gauges. In order to make contact with previous works, we will 
use the notation of Ref. \cite{G1}. There the quantity ${\cal H}$ is 
denoted by $\alpha $, $\alpha \equiv {\cal H} \equiv a'/a$ and the 
function $\gamma(\eta )$ is defined by: 
\begin{equation}
\label{2-17}
\gamma \equiv 1- \frac{\alpha'}{\alpha ^2} \ .
\end{equation}
This function reduces to a constant for scale factors which are
proportional to a power of the conformal time.
The function $\gamma$ is zero if
$\varphi _0'=0$ (de Sitter). In Ref. \cite{G1}, the line element 
is written as:
\begin{equation}
\label{ds2Gri}
{\rm d}s^2=a^2(\eta )\{-{\rm d}\eta ^2+[(1+hQ)\gamma _{ij}+\frac{h_l}{k^2-K}
Q_{|i|j}]{\rm d}x^i{\rm d}x^j\}.
\end{equation}
Choosing the synchronous gauge (SG) means setting the perturbed lapse 
and shift functions to zero. The notation of \cite{G1} is related 
to the notation of \cite{MFB} by
\begin{equation}
\label{gitosyn}
\phi =0, \qquad B=0, \qquad \psi =-\frac{h}{2}Q, \qquad 
E=\frac{h_l}{2(k^2- K)}Q \ .
\end{equation}
Thus, the gauge-invariant variables expressed in the synchronous 
gauge are given by:
\begin{eqnarray}
\label{2-20}
\Psi ^{(SG)} &=& -\frac{1}{2}(h-\frac{\alpha }{k^2-K}h_l') \ , \\
\label{2-21}
\Phi ^{(SG)} &=& -\frac{1}{2(k^2-K)}(h_l''+\alpha h_l') \ , \\
\label{2-22}
\delta \varphi ^{(gi)(SG)} &=& 
(\varphi _1-\frac{1}{2(k^2-K)}\varphi _0'h_l') \ ,
\end{eqnarray}
where the scalar field is written as: $\varphi \equiv \varphi _0(\eta )+
\varphi _1(\eta )Q$. Inserting these formulas into the system 
(\ref{2-11}) -- (\ref{2-14}) provides the correct perturbed Einstein 
equations in the synchronous gauge (see Ref. \cite{G1}), namely:
\begin{eqnarray}
\label{2-23}
h_l''+2\alpha h_l'-(k^2-K)h &=& 0 \ , \\
\label{2-24}
-h' +\frac{K}{k^2-K}h_l &=& \kappa \varphi _0'\varphi _1 \ , \\
\label{2-25}
3\alpha h'-\alpha h_l'+(k^2-4K)h &=& 
\kappa (\varphi _0'\varphi _1'+
a^2\varphi _1 V_{,\varphi}) \ , \\
\label{2-26}
-h''-2\alpha h'+Kh &=& \kappa (\varphi _0'\varphi _1'-
a^2\varphi _1 V_{,\varphi}) \ .
\end{eqnarray}
This shows that, as expected, the two formalisms are completely equivalent. 
It is worth noticing that equation (\ref{2-23}) expresses the fact that 
$\Psi =\Phi $. This is due to the vanishing longitudinal (anisotropic)
pressure (see Ref. \cite{G1}). 

Let us now consider the question of the residual gauge. It has been 
known for a long time that the condition $h_{0\mu }=0$ does not fix 
the gauge completely. This condition is preserved under the 
change of coordinates
\begin{equation}
\label{2-28}
\bar{\eta }=\eta -\frac{C}{2a}Q, \qquad \bar{x}^i=x^i-
\frac{C}{2}Q^{|i}\int \frac{{\rm d}\eta }{a}-
\frac{D}{2}Q^{|i} \ ,
\end{equation}
where $C$ and $D$ are arbitrary constants for a fixed mode $k$. 
The corresponding changes for $h$ and $h_l$ are given by:
\begin{equation}
\bar{h}=h+C\frac{\alpha }{a}, \qquad \bar{h}_l=h_l+
k^2C\int \frac{{\rm d}\eta }{a}+k^2D \ .
\end{equation}
According to Ref. \cite{G1}, one can thus introduce two 
`residual-gauge-invariant' quantities $u$ and $v$ defined by:
\begin{equation}
\label{2-30}
u\equiv h'+\alpha \gamma h, \qquad v\equiv h_l'-\frac{k^2}{\alpha }h \ .
\end{equation}
The relations between the gauge-invariant quantities and the 
residual-gauge-invariant quantities can be expressed as:
\begin{eqnarray}
\label{2-31}
\Psi ^{(SG)} &=& \frac{\alpha }{2k^2}v \ , \\
\label{2-32}
\Phi ^{(SG)} &=& -\frac{u}{2\alpha }-\frac{1}{2k^2}(v'+\alpha v) \ .
\end{eqnarray}
In Ref. \cite{G1} it has been claimed that $u$ and $v$ are genuine 
gauge-invariant quantities [i.e., invariant under the transformation 
(\ref{2-6})]. This claim is not correct. Indeed, a direct check shows that:
\begin{eqnarray}
\label{2-33}
\bar{u} &=& u+2\alpha \xi ^{0'}+2\alpha ^2\xi ^0 \ , \\
\label{2-34}
\bar{v} &=& v+2k^2(\xi '-\xi ^0) \ .
\end{eqnarray}
The equations (\ref{2-31}) -- (\ref{2-32}) just give the value $\Psi $ and 
$\Phi $ calculated in the synchronous gauge, $\Psi ^{(SG)}$, $\Phi ^{(SG)}$. 
It does not come as a surprise that what remains in a fixed gauge from the 
gauge invariance is simply the residual gauge invariance. 

In Refs. \cite{G2} and \cite{G3}, it has been argued that equation 
(\ref{2-15}) is incorrect since it could be expressed as a combination of 
the derivatives of the correct equations and hence would contain a
non-physical constant. This claim is incorrect as well. Let us demonstrate 
why. If we insert the expression of $\Phi $ in terms of $u$ and $v$ 
[equation (\ref{2-32})] in equation (\ref{2-15}), we find that this 
one transforms identically into:
\begin{eqnarray}
\label{35}
& & \frac{1}{2k^2\alpha }[k^2u+\alpha (v'+2\alpha v)]''+
\frac{1}{k^2}(2\frac{\alpha '}{\alpha ^2}+\frac{\gamma '}{2\alpha \gamma }-
\frac{1}{2})[k^2u+\alpha (v'+2\alpha v)]' \nonumber \\
& & +(\frac{\alpha '}{2k^2\alpha }-\frac{2\alpha ^{'2}}{k^2\alpha ^3}-
\frac{\alpha '\gamma '}{2k^2\alpha ^2\gamma }+\frac{\alpha ''}{2k^2\alpha ^2}-
\frac{1}{2\alpha })[k^2u+\alpha (v'+2\alpha v)] \nonumber \\
& & -\frac{1}{2}[u'-\alpha v-(\frac{\alpha '}{\alpha }+
\frac{\gamma '}{\gamma })u] =0 \ .
\end{eqnarray}
Of course we must also take into account the equation $\Psi =\Phi $ 
which is, in terms of $u$ and $v$, equal to: $-k^2u=\alpha (v'+2\alpha v)$. 
Therefore, we see that from the gauge-invariant formalism, we can 
reduce the whole problem to a set of two coupled first order 
differential equations (of course, this could have been done 
directly in the synchronous gauge) for the variables $u$ and $v$:
\begin{eqnarray}
\label{2-36}
-k^2u &=& \alpha (v'+2\alpha v) \ , \\
\label{2-37}
v &=& \gamma (\frac{u}{\alpha \gamma })' \ .
\end{eqnarray}
{}From this system, we can generate two decoupled second order 
differential equations:
\begin{eqnarray}
\label{2-38}
u''+u'(2\alpha \gamma -\frac{\gamma '}{\gamma })+u[k^2-2\alpha '-
\alpha \frac{\gamma '}{\gamma }-(\frac{\gamma '}{\gamma })'] &=& 0 \ , \\
\label{2-39}
v''+v'(2\alpha -\frac{\gamma '}{\gamma })+k^2v+v(2\alpha '-
2\alpha \frac{\gamma '}{\gamma }) &=& 0 \ .
\end{eqnarray}
The last equation could have been guessed from the very beginning 
by inserting the expression $\Phi^{(SG)}=\Psi^{(SG)}=(\alpha v)/(2k^2)$ 
in 
equation (\ref{2-15}). The variable change (see Ref. \cite{G1}) 
\begin{equation}
\label{mu}
u \equiv {\alpha \sqrt{\gamma }\over a} \mu 
\end{equation}
in formula (\ref{2-38}) allows us to obtain the correct equation for $\mu$:
\begin{equation}
\label{2-40}
\mu ''+[k^2-\frac{(a\sqrt{\gamma })''}{a\sqrt{\gamma }}]\mu =0 \ .
\end{equation}
Therefore, we have proved that the 
gauge-invariant framework leads to the same (correct) equation of motion 
for the variable $\mu$ as the calculation in synchronous gauge. Let us 
note that the residual-gauge-invariant variable $\mu$ is nothing but the 
value of the gauge-invariant variable 
\begin{equation}
\label{vM}
v_{\rm M}\equiv a (\delta \varphi^{(gi)} + {\varphi_0'\over {\cal H}} \Phi )\ ,
\end{equation}
which was defined by Mukhanov in Refs.~\cite{M88,MFB}, expressed 
in the synchronous gauge. Inserting the synchronous gauge values 
of $\delta \varphi^{(gi)}$ and $\Phi$ into (\ref{vM}) and 
using (\ref{2-36}) yields:
\begin{equation}
\label{vMmu}
v^{\rm (SG)}_{\rm M} = - {\mu\over\sqrt{2 \kappa}} \ .
\end{equation}
   
In order to be as complete as 
possible we examine what was the mistake of Refs.~\cite{G2,G3}. 
{}From the equation (\ref{2-31}) and the definition of $\mu$, we can 
re-write equation (\ref{2-37}) as:
\begin{equation}
\label{2-41}
\Phi ^{(SG)}=\frac{\alpha \gamma }{2k^2}(\frac{\mu }{a\sqrt{\gamma }})' \ .
\end{equation}
If one inserts this expression in formula (\ref{2-15}), we obtain 
a third order differential equation for $\mu $:
\begin{equation}
\label{2-42}
\{\mu ''+[k^2-\frac{(a\sqrt{\gamma })''}{a\sqrt{\gamma }}]\mu \}'-
\frac{(a\sqrt{\gamma })'}{a\sqrt{\gamma }}\{\mu ''+
[k^2-\frac{(a\sqrt{\gamma })''}{a\sqrt{\gamma }}]\mu \}=0 \ ,
\end{equation}
which can be integrated and leads to:
\begin{equation}
\label{2-43}
\mu ''+ [k^2-\frac{(a\sqrt{\gamma })''}{a\sqrt{\gamma }}] \mu =
Xa\sqrt{\gamma } \ ,
\end{equation}
where $X$ is a constant of integration. Comparing equations (\ref{2-40}) 
and (\ref{2-43}) shows that $X=0$.

Let us demonstrate why the introduction of $\mu$ via (\ref{2-41}) and
equation (\ref{2-15}) seems to give $X\neq0$ at first sight.
Consider a set of two first order differential equations [they 
play the role of equations (\ref{2-36}) and (\ref{2-37}), the latter is 
equivalent to (\ref{2-41})]: $x'=-y,\ y'=x$.
{}From them we can generate two second order differential equations
[they play the role of equations (\ref{2-38}) and (\ref{2-39}), which is 
nothing but equation (\ref{2-15})]: 
$y''+y=0$ and $x''+x=0$. But we can also insert $y'=x$ 
into the second of the two last equations [as we inserted (\ref{2-41}) 
into (\ref{2-15})]. We will obtain a third order differential equation, 
$y'''+y'=0$, which may be integrated to yield $y''+y=X$. There is 
no harm to do that as long as we do not forget to use  
equation $y''+y=0$ and therefore $X$ must be equal to zero.

However, in Refs.~\cite{G2,G3} the derivation of equation (\ref{2-41}) 
was not given, instead it was assumed that (\ref{2-41}) may be
considered to be the definition of $\mu$. In that case, 
this 'new' $\mu$ has nothing 
to do with the $\mu$ defined previously and has not to satisfy 
Eq.~(\ref{2-40}). The equation 
of motion for this 'new' variable $\mu $ is Eq. (\ref{2-43}). 
This definition of $\mu$ is unique up to a shift $\mu \to 
\mu + Y a\sqrt{\gamma}$ only, where $Y$ is an arbitrary constant for a fixed 
mode $k$. Demanding that $\mu$ should fulfill (\ref{2-40})
fixes $Y = X/k^2$. The choice $Y=0, X\neq 0$ that implicitly 
was made in~\cite{G2,G3} is inconsistent. 

In conclusion, let us emphasize again the main result of this section. 
The gauge-invariant formalism and the synchronous gauge formalism are 
completely equivalent for all values of $k$ including the zero-mode. Using 
one or the other (or any further gauge) is only a question of taste or 
of prejudices. 

\section{The constancy of $\zeta$ for superhorizon modes}

In this section we turn to the study of the so-called `conservation law'. Let
us start by introducing matter that can be described by a hydrodynamical 
equation of state, in order to follow the evolution of the perturbations from
reheating till today.

\subsection{Perfect fluids and the `standard result'}

Assume that anisotropic stresses can be neglected, thus matter is described
by a perfect fluid. The pressure of the perfect fluid
is related to its energy density by the equation of state $p = p(\rho,S)$.
$S$ is the entropy per baryon\footnote{Of course, this makes sense  
when baryon number is conserved only.}, $n$ is the density of 
baryons. For a reversible expansion of the background (there are no 
unbalanced creation/annihilation processes) the entropy per baryon is 
constant in time. Due to the isotropy of the background, the first law of
thermodynamics reads ${\rm d}(\rho/n) = - p{\rm d}(1/n)$, thus 
$T{\rm d}S = 0$ from the 
second law for reversible processes. Therefore,
the background equation of state has to be isentropic ($\nabla_i S =0$),
thus we may write $p = p(\rho) \equiv w(\rho) \rho$ for an adiabatically 
expanding background.

The pressure perturbation reads $\delta p = c_s^2 \delta \rho + \tau \delta S$, 
where $c_s^2 \equiv (\partial p/\partial\rho)_s$ is the isentropic sound speed
and $\tau \equiv (\partial p/\partial S)_\rho$. 
For an adiabatically expanding background $p' = c_s^2 \rho'$. 
Below we use the relation $c_s^2 = p'/\rho'$, thus the following results 
hold true for negligible entropy production only.

The equations of motion for the gauge-invariant metric potentials for 
a perfect fluid are (see Ref.~\cite{MFB}) $\Phi = \Psi$ and 
\begin{equation}
\label{eompf}
\Phi'' + 3(1 + c_s^2){\cal H}\Phi' + [2{\cal H}' + (1 + 3 c_s^2)({\cal H}^2 
- K)] \Phi + c_s^2 k^2 \Phi = \frac{\kappa}{2} a^2 \tau \delta S \ ,
\end{equation}
except for the de Sitter universe, where $\Phi = 0$.
Written in terms of the variable $\sigma$, which is defined by
(\ref{sigma}), the latter equation reads\footnote{With the definition
$\theta \equiv 1/a [\rho_0/(\rho_0+p_0)]^{1/2}(1 - 3K/\kappa \rho_0 a^2)^{1/2}$
Eq.~(\ref{spf}) holds true for all values of $K$.}:
\begin{equation}
\label{spf}
\sigma'' +\left(c_s^2 k^2 - {\theta''\over \theta}\right)
\sigma = {\kappa\over 3} {\theta\over {\cal H}} a^4 \tau\delta S \ .
\end{equation}
For isentropic perturbations the leading order solution is easily obtained 
to be $\sigma = \tilde{C}_1 \theta \int {\rm d}\eta/\theta^2 + 
\tilde{C}_2 \theta$, whereas the 
next to leading terms differ from the solution (\ref{sigmaexp}), because
$c_s^2$ may be time dependent. [There should be a factor $c_s^2$ in the 
$\bar{\eta}$ integration in (\ref{sigmaexp}).] If $w$ and the sound speed
are constant an approximate solution of $\sigma$ is not of much use, 
because then the exact solution
to all orders in $k$ can be given in terms of Bessel functions.
Two examples of a perfect fluid are
the radiation fluid ($w = c_s^2=1/3$) and dust ($w = c_s^2 = 0$).
These examples have vanishing entropy perturbations ($\delta S = 0$).
In general, $\delta S$ does not vanish for more than one fluid. 
The scalar field $\varphi$ is another form of matter that can be described by
a perfect fluid, see~(\ref{2-2}). Formally, the sound speed is defined as 
above. With help of the Klein-Gordon equation (\ref{2-3}) it reads
\begin{equation}
\label{csphi}
c_s^2(\varphi_0) = -\frac13 \left( 1 + {2 \varphi_0''\over {\cal H}
\varphi_0'}\right) \ .
\end{equation}
Now we define the `entropy perturbation' through 
$\tau \delta S = \delta p - c_s^2 \delta \rho$. 
With the expressions for $\delta\rho$ and $\delta p$ and 
using Eqs.~(\ref{2-11}) and (\ref{2-12}) we arrive at
\begin{equation}
\label{deltaS}
\frac{\kappa}{2} a^2 \tau \delta S = \left(1 - c_s^2(\varphi_0)\right)
\left( 3 K - k^2 \right) \Phi \ .
\end{equation}
We obtain the scalar-field equation of motion for the metric potential, 
Eq.~(\ref{2-15}), by inserting (\ref{deltaS}) into (\ref{eompf}) and 
replacing $c_s^2$ by (\ref{csphi}). Thus, we may study scalar metric
perturbations by means of Eq.~(\ref{eompf}) 
from their generation during the inflation epoch to its observation
in the CMBR today.
However, during reheating \cite{reheating} the scalar field 
may oscillate, thus $\varphi_0'$ has zeros and Eq. (\ref{2-15}) is singular 
at these points. This situation has been discussed in \cite{Kodama}. Let us 
assume in this work that this does not happen.

We are now able to derive the `standard result', i.e. the amplification of
$\Phi$ during the reheating transition. From comparison of (\ref{sigmaexp})
with the solution of (\ref{spf}) it is clear that the leading superhorizon
term of the solution does not depend on the sound speed $c_s$. After some
time the decaying mode is unimportant and the leading superhorizon 
term, using (\ref{sigma}), reads:
\begin{equation}
\label{sigmagrm}
\Phi \simeq \frac32 \tilde{A}_1^(k) {{\cal H}\over a^2} 
\int^\eta a^2 (1+w) {\rm d}\bar{\eta} = \tilde{A}_1(k) 
{{\cal H}\over a^2} \int^\eta a^2 \left(1 - {{\cal H}'\over {\cal H}^2}
\right) {\rm d}\bar{\eta} \ .
\end{equation}
For a power law behavior of the scale factor, i.e. 
$a \propto |\eta|^{1+\beta}$, the equation of state is given by
$w = (1-\beta)/[3(1+\beta)]$. Then, the `growing' mode is constant in time: 
\begin{equation}
\Phi \simeq \frac32 \tilde{A}_1(k) {1+\beta\over 2\beta + 3} 
(1 + w) \ . 
\end{equation}
Let us assume that the evolution of the scale factor may be described by
such a power law far away from the transitions inflation-radiation and 
radiation-matter. In between it changes smoothly. 
The ratio of the values of $\Phi$ 
during inflation and matter domination (again far away from the transitions)
is then given by
\begin{equation}
\label{sr}
{\Phi_m \over \Phi_i} \simeq \frac25 
{2\beta_i + 3\over \beta_i + 1} {1\over 1+w_i} \approx
\frac25 {1\over 1+w_i} \ ,
\end{equation}
where $\beta_m = 1, \beta_i \approx -2$, and $w_i \approx -1$.
Therefore, scalar perturbations are magnified by a big factor during the
reheating transition.
For de Sitter spacetime the amplification coefficient 
goes to infinity. This simply expresses the trivial fact that $\Phi$
goes from zero to a constant without prejudice to the numerical 
value of this constant. Note that we can not conclude from this argument that
$\Phi_m$ is large because $\Phi_i \to 0$ as $w_i \to -1$!
The real (absolute) value of $\Phi$ after 
the transition can be only known after having determined the initial 
conditions from the quantization of density fluctuations, that is to say 
after having fixed $\tilde{A}_1$.

\subsection{Definitions and use of $\zeta$}

As was first recognized by Bardeen, Steinhardt and Turner \cite{BST}
and further elaborated in \cite{FT,BK,L,B2}, the equation of motion 
(\ref{eompf}) has
a first integral for isentropic modes ($\delta S = 0$) that are 
much larger than the Hubble scale, i.e.
$k^{\rm phys} \equiv k/a \ll H$. Following Ref.~\cite{MFB} we define 
\begin{equation}
\label{defzeta}
\zeta \equiv \frac 23 {{\cal H}^{-1} \Phi^\prime + \Phi\over 1 + w} + \Phi \ ,
\end{equation}
which was introduced by Lyth \cite{L} originally (a quantity differing by
terms ${\cal O}(k^2/{\cal H}^2)$ only 
was used by Brandenberger and Kahn \cite{BK}). 
$\zeta$ essentially is the perturbation of 
the intrinsic curvature in the comoving gauge \cite{L}. 
Its first derivative reads 
\begin{equation}
\label{zp}
{1\over {\cal H}} \zeta^\prime = \frac23 {1 \over (1+w)} 
\left[{K\over {\cal H}^2} \left({1+3w\over 2{\cal H}}
\Phi^\prime + 3(c_s^2 - w) \Phi \right) + 
\frac{\kappa}{2}{a^2 \tau\delta S\over {\cal H}^2}
- c_s^2 \left(k\over {\cal H}\right)^2 \Phi \right] \ ,
\end{equation}
where we used the equation of motion (\ref{eompf}) and the equations
\begin{equation}
w' = 3{\cal H}(1+w)(w - c_s^2) \ , \qquad 
{\cal H}' = -\frac{1+3w}{2}({\cal H}^2 + K) \ .
\end{equation}
Let us note that (\ref{zp}) is an equation for a fixed mode $k$, i.e.
the large scale limit $k/{\cal H} \to 0$ in this equations means to make 
${\cal H}$ large.   
Thus, $\zeta$ is constant in time 
for superhorizon modes $k/{\cal H} \ll 1$ if and only
(i) $K = 0$, (ii) entropy perturbations are negligible, and (iii) 
the perturbation is given by the regular mode only. 
The conditions (i) and (ii) are obvious from (\ref{zp}).
The last condition means that the last term on the r.h.s of (\ref{zp}) 
vanishes in 
the limit $k/{\cal H} \to 0$. The decaying mode is $\Phi \propto
{\cal H} \sigma /(a^2 \theta) \propto {\cal H}/a^2$ at leading order.
Thus, the r.h.s
is proportional to $(k/{\cal H})^2 {\cal H}/a^2$, which blows up in
the limit $a \to 0$, except for very special cases\footnote{
In the model given by the scale factor $a \propto \eta^{1+\beta}$ with
$-1 < \beta < -1/2$ and $\eta >0$ the decaying mode does no harm, because
$(k/{\cal H})^2 {\cal H}/a^2 \propto \eta^{-2\beta-1}$. However, these
models lead to equations of state where $3 p_0 > \rho_0$, which does not
give rise to inflationary expansion.}. 
Therefore, in general, one has to exclude the 
decaying mode in order to make use
of the `conservation law' $\zeta' = 0$. In fact $\zeta$ is nothing more
than a first integral of the equation of motion (\ref{eompf}) when the
conditions (i) -- (iii) are fulfilled.

The above definition of $\zeta$ differs from the original definition 
\cite{BST,B2} 
\begin{equation}
\zeta_{\rm BST} \equiv \frac13 {\delta\rho\over \rho_0 + p_0} - \psi \ .
\end{equation}
$\zeta_{\rm BST}$ is a hypersurface-independent quantity \cite{B2}.
Written in terms of the gauge-invariant metric potential it reads:
\begin{equation}
\zeta_{\rm BST} = - \frac 23 {{\cal H}^2 \over (1 + w) ({\cal H}^2 + K)}
\left[{\cal H}^{-1} \Phi^\prime + (1 - {K\over {\cal H}^2} + \frac13 
\left(k\over {\cal H}\right)^2)\Phi\right] - 
\Phi \ .
\end{equation}
{}From its time derivative 
\begin{equation}
{1\over {\cal H}} \zeta_{\rm BST}' = 
- \frac23 {{\cal H}^2\over (1+w)({\cal H}^2 + K)}
\left[ \frac13 \left(k\over {\cal H}\right)^2 ({\cal H}^{-1} \Phi' + \Phi)
+\frac{\kappa}{2} {a^2 \tau \delta S\over {\cal H}^2} \right] 
\end{equation}
the constancy of $\zeta_{\rm BST}$ follows if and only (i) there are no entropy
perturbations, and (ii) there is the regular mode only. 

Besides the advantage 
of $\zeta_{\rm BST}$ over $\zeta$ to be conserved even if $K \neq 0$, 
$\zeta_{\rm BST}$ is a hypersurface-independent measure of the metric 
perturbations including regular and singular modes 
(singular in the limit $k/{\cal H} \to 0$), whereas $\zeta$ does not measure 
singular modes, if one takes the leading order contribution into
account only. This can be easily seen by rewriting $\zeta$ in terms of
$\sigma$: Whatever the coefficient $\tilde{C}_2$ in front of the 
decaying mode
is, it does not enter into $\zeta$ in the 
leading order in $k/{\cal H}$, because
\begin{equation}
\label{zetatheta}
\zeta = \theta^2 \left(\sigma\over\theta\right)' \ ,
\end{equation}
for $K = 0$.  On the other hand
\begin{equation}
\zeta_{\rm BST} = - \theta \left[\sigma' -
({\theta'\over\theta} - \frac 13 {k^2\over {\cal H}} )\sigma \right] \ .
\end{equation}
(for any value of $K$) does depend on both $\tilde{C}_1$ and 
$\tilde{C}_2$. Let us 
in the following discuss the properties of $\zeta(k)$.

In Ref.~\cite{G2} it has been claimed that ``the conservation law
$\zeta(t_i) = \zeta(t_f)$ degenerates to an empty statement $0=0$''.
In order to explain the line of reasoning
of \cite{G2} and to show where the argument fails, 
let us restrict the discussion to matter in form of a scalar field.
For an isentropic perfect fluid the line of reasoning is analogous. 
Since $\zeta$ is a gauge-invariant variable its value is the same in 
all gauges. Let us compute it in the synchronous gauge. Inserting 
Eq. (\ref{2-41}) in Eq. (\ref{defzeta}) we find that:
\begin{equation}
\label{zetamu}
\zeta=\frac{1}{2k^2a^2\gamma }[a^2\gamma (\frac{\mu }{a
\sqrt{\gamma }})']'.
\end{equation}
In Ref.~\cite{G2} Eq. (\ref{2-43}), re-written as 
\begin{equation}
\label{eqofmX}
\frac{1}{a^2\gamma }[a^2\gamma (\frac{\mu }{a\sqrt{\gamma }})']'
+k^2\frac{\mu }{a\sqrt{\gamma }}=X,
\end{equation}
was inserted into (\ref{zetamu}) to obtain $\zeta\sim X/(2k^2)$ 
in the limit $k\rightarrow 0$. According to Ref. \cite{G2}, 
there is no mean to know that 
$X=0$ in the gauge-invariant formalism. This would explain  
$\zeta = const. \neq 0$ in the considered limit. On the 
other hand, always according to Ref. \cite{G2}, the synchronous 
gauge formalism could tell us that $X=0$. Thus, we would be `betrayed' 
by the gauge-invariant formalism in which $X$ would 
appear. Only computations in the synchronous gauge formalism could 
reveal that $\zeta=X/(2k^2)=0$. In Refs.~\cite{G2,G3},
the fact that $\zeta=0$ at the `leading order'
$k^{-2}$ was misused as a proof that it is not possible 
to use the quantity $\zeta$ to learn something about 
the behavior of density perturbations.

We have shown in Sec.~III that the synchronous gauge formalism 
and the gauge-invariant formalism are completely equivalent and 
that $X=0$ in both approaches. Thus, there is no risk of confusion at 
all computing $\zeta$ in the gauge-invariant formalism. 

According to Ref. \cite{C} (p.~6), the conclusion of Ref.~\cite{G2} 
occurs because ``the $k\rightarrow 0$ limit has not been 
taken consistently''. Let us show that this is not the case.
Following Ref.~\cite{G2} we assume $X\neq0$ because 
we now define $\mu$ by Eq.~(\ref{2-41}), but remember that this definition 
of $\mu$ is not unique. In that case the equation of motion is given by 
Eq. (\ref{2-43}). Let us expand the solution $\mu(\eta)$ in powers of $k^2$:
\begin{eqnarray}
\label{expmures}
{\mu\over a\sqrt{\gamma}} &=& 
\bar{A}_1(k) \left(1 - k^2 \int\frac{1}{a^2\gamma}\int^\eta a^2\gamma 
{\rm d}\bar{\eta} {\rm d}\eta + {\cal O}(k^4) \right) + \nonumber \\
& & + \bar{A}_2(k) \int\frac{1}{a^2\gamma}\left(
1 - k^2 \int^\eta {a^2\gamma} \int^{\bar{\eta}} \frac{1}{a^2\gamma} 
{\rm d}\tilde{\eta} {\rm d}\bar{\eta} +{\cal O}(k^4)
\right){\rm d}\eta + \\
& & + X \int\frac{1}{a^2\gamma}\int^\eta a^2\gamma \left( 1 - k^2 
\int^{\bar{\eta}}\frac{1}{a^2\gamma}\int^{\tilde{\eta}} a^2\gamma 
{\rm d}\hat{\eta} {\rm d}\tilde{\eta} 
 + {\cal O}(k^4)\right) {\rm d}\bar{\eta} {\rm d}\eta
\nonumber \ ,
\end{eqnarray}
where
$\bar{A}_1(k)$, $\bar{A}_2(k)$ are arbitrary integration 
constants fixed 
by the initial conditions. If we compare Eq. (\ref{zetamu}) with 
Eq. (\ref{eqofmX}) we deduce that:
\begin{equation}
\label{zetaX}
\zeta=\frac{X}{2k^2}-\frac{\mu }{2a\sqrt{\gamma }} \ ,
\end{equation}
and we find the beginning of the series giving $\zeta$:
\begin{eqnarray}
\label{expzetaX}
\zeta &=& \frac{X}{2k^2}-\frac{1}{2}\biggl(\bar{A}_1(k)
+ \bar{A}_2(k)\int \frac{{\rm d}\eta }{a^2\gamma} 
+ X \int \frac{1}{a^2\gamma}\int^\eta a^2\gamma {\rm d}\bar{\eta} {\rm d}\eta  
+ {\cal O}(k^2) \biggr) \ .
\end{eqnarray}
This confirms that the limit had been taken properly in Ref.~\cite{G2} and
that the first term in the series contains $X/k^2$. 
This result does not imply $\zeta \to \infty$ as $k\to 0$,
because $X = X(k)$ and we know nothing about the $k$ dependence of $X$. 
However, there is no argument telling us that $\bar{A}_1(k)$ or 
$\bar{A}_2(k)$ are subleading compared to $X(k)/k^2$.
\par
Taking into account the correct value $X=0$, Eq.~(\ref{expzetaX})
reads:
\begin{equation}
\label{expzeta}
\zeta=-\frac{\bar{A}_1}{2}-\frac{\bar{A}_2}{2}\int 
\frac{{\rm d}\eta }{(a\sqrt{\gamma })^2}+{\cal O}(k^2).
\end{equation}
The same expression could have been obtained directly within 
the gauge-invariant formalism. It is necessary to push the 
expansion in $k^2$ one step further because the leading order term of the 
decaying mode does not contribute in (\ref{zetatheta}). 
Inserting Eq.~(\ref{sigmaexp}) in Eq.~(\ref{zetatheta}), we obtain:
\begin{equation}
\label{expzetatheta}
\zeta= \tilde{A}_1 - \tilde{A}_2 k^2 
\int{\rm d}\eta \theta^2 +{\cal O}(k^2) \ .
\end{equation}
Comparison with (\ref{expzeta}) yields 
$\bar{A}_1(k) = -2\tilde{A}_1(k)$ and 
$\bar{A}_2(k) = 2k^2\tilde{A}_2(k)$. We would have obtained 
the same relations between 
the initial conditions by inserting (\ref{expmures}) into (\ref{2-41}) 
and comparing the result with (\ref{sigmaexp}). 
{}From (\ref{expzetatheta}) it is seen that $\zeta$ does not vanish and is not 
constant in general. The second term of (\ref{expzetatheta}) 
is the decaying mode. We emphasize that it is crucial 
to neglect the decaying mode if one makes use of  
\begin{equation}
\label{approxzeta}
\zeta \simeq \tilde{A}_1 \ .
\end{equation}

Let us show how to make use of the constancy of $\zeta$. Assume 
conditions (i) -- (iii) are fulfilled. An example is the `standard'
scenario: The initial perturbations are provided by quantum-fluctuations
during inflation. At the first horizon crossing we denote them $\Phi_k(t_i)$.
Let $k$ be a mode which re-enters the Hubble horizon after equality between
matter and radiation. The theory of quantum fluctuations shows that
$\Phi_k(t_i)$ is due to isentropic (adiabatic) fluctuations. The 
decaying mode is negligible short after the first horizon crossing. 
Entropy perturbations due to the reheating transition and due to the 
transition from radiation to matter affect much smaller scales.
We know that the growing mode in $\Phi$ is a constant on superhorizon scales
if the scale factor obeys a simple power law behavior. In the inflationary 
stage $a \propto t^p$ where $p\gg 1$, in the matter stage $a \propto t^{3/2}$.
Thus, with $w(t_m) \approx 0$  
\begin{equation}
\zeta \simeq {\frac53 + w(t_i)\over 1 + w(t_i)} \Phi_k(t_i) \simeq
              \frac53 \Phi_k(t_m) \ .
\end{equation} 
During inflation $w \sim -1$ and therefore the large amplification
\begin{equation}
\Phi_k(t_m) \simeq \frac35 \zeta \simeq \frac25 {1\over 1 + w(t_i)} 
\Phi_k(t_i)
\end{equation}
follows, which is the same as (\ref{sr}). 

To finish this section let us consider the concrete model studied 
in Ref. \cite{G3} (see the appendix of that paper). It consists in a 
transition from one power law scale factor, $a(t)=a_1t^{p_1}$ for $t<t_1$, 
to another power law scale factor, $a=a_2(t-t_*)^{p_2}$ for $t>t_1$. At 
the transition $a$ and $H$ are continuous, whereas $\dot{H}$ jumps. 
Assuming the constancy of $\zeta=\zeta_0$ we may integrate the definition of
$\zeta$ (\ref{defzeta}) in terms of cosmic time 
to obtain the evolution of the Bardeen potential
\begin{equation}
\label{Phit}
\Phi(t)= -\zeta_0 \frac{H}{a}\int_{t_i}\frac{a\dot{H}}{H^2}{\rm d}t + 
\frac{H}{a} C\ .
\end{equation}
For $t<t_1$ we find: 
\begin{equation}
\label{Phiinf}
\Phi (t<t_1)=\frac{\zeta_0}{1+p_1}[1-(\frac{t_i}{t})^{1+p_1}] + 
\frac{p_1}{a_1 t^{1+p_1}} C \ .
\end{equation}
Consistent use of the constancy of $\zeta =\zeta_0$ 
requires the vanishing of the decaying mode in Eq. (\ref{Phiinf}) and 
therefore $C = \zeta_0 a_1 t_i^{1+p_1}/(p_1+p_1^2)$. We obtain:
\begin{equation}
\label{Phiinfconst}
\Phi (t<t_1)=\frac{\zeta_0}{1+p_1}.
\end{equation}
For the de Sitter case (`$p_1=\infty $') we recover that $\Phi =0$. Let 
us compare this result with the one of Ref. \cite{G3}. We do not 
agree that ``The initial value of the potential is $\Phi(t_i) = 
\zeta_0 + H(t_i)C/a(t_i)$, \dots'' and that 
``The constant $C$ could be set to zero from the very beginning.'' 
as it is stated on page 31. This would lead to $\Phi(t_i)=\zeta_0$ 
and, would imply a non-vanishing Bardeen's potential 
for the de Sitter spacetime. Instead the initial value is $\Phi(t_i) = 
\zeta_0/(1+p_1)$ which is clear from Eq.~(\ref{Phiinfconst}).

For $t>t_1$, we obtain the solution:
\begin{equation}
\label{Phisup}
\Phi (t>t_1)=\frac{\zeta_0}{1+p_1}(\frac{p_2}{p_1})^{1+p_2}
(\frac{t_1}{t-t_*})^{1+p_2}+\frac{\zeta_0}{1+p_2}[1-
(\frac{p_2}{p_1})^{1+p_2}(\frac{t_1}{t-t_*})^{1+p_2}].
\end{equation}
The Bardeen variable is continuous at $t=t_1$ (since we have 
integrated a Heaviside function). Long after the transition, 
the previous relation reduces to:
\begin{equation}
\label{Phiverysup}
\Phi(t\gg t_1)\simeq\frac{\zeta_0}{1+p_2}.
\end{equation}
Therefore, we reach the conclusion that the amplification coefficient is 
given by:
\begin{equation}
\label{amplicoef}
\frac{\Phi (t \gg t_1)}{\Phi (t<t_1)}\simeq \frac{1+p_1}{1+p_2}.
\end{equation}
Formula (\ref{amplicoef}) should be compared with the third equation 
after Eq. (87) of Ref. \cite{G3}. In this expression the missing term 
$1+p_1$ in the numerator is due to the incorrect assumption 
$\Phi(t_i)=\zeta_0$. 

\section{A sharp transition --- Joining conditions}

There are two physical situations where a sharp transition in the equation
of state is a good approximation. The first one is the 
study of the behavior of superhorizon modes ($k^{\rm phys} \ll H$). The 
duration $T$ of processes
like the change from the radiation dominated to the matter dominated universe,
reheating at the end of inflation, or recombination typically take several
expansion times, i.e. $T \sim H^{-1}$. Superhorizon modes change on much
larger time scales, i.e. $1/k^{\rm phys}$, thus they see a sharp transition.
These sharp transitions may violate the second law of thermodynamics,
e.g. at equality radiation entropy is destroyed instantaneously
when we assume a sharp
drop in the pressure. The second situation where sharp transitions are 
of interest in cosmology are phase transitions like the QCD transition or a 
GUT transition. In these transitions the pressure is continuous, but its 
derivatives may be discontinuous. An example is provided by a first order QCD 
transition where the sound velocity may jump \cite{SSW}. 

Recently, Deruelle and Mukhanov \cite{DM}
derived the joining conditions for
scalar metric perturbations in a spatially flat FLRW model. 
For perfect fluids (no scalar fields) gauge-invariant joining conditions
for cosmological perturbations have been derived before by Hwang and 
Vishniac \cite{Hwang}. The difficulty to state the correct joining conditions 
arises because the physical hypersurface of the 
transition is not necessarily that of constant coordinate time. 
In the equality and reheating transitions the physical hypersurface 
is the one of constant density contrast.
The gauge-invariant variables describe zero shear hypersurfaces as constant
time hypersurfaces. Below we derive
the joining conditions for general, spatially non-flat metric perturbations. 
Although the method of Deruelle and Mukhanov might be simpler than ours,
we think that it is worth to view the problem from a different perspective
below.
\par
Let us start with exposing the method in general. Assume the
spatial transition hypersurface $\Sigma$ is defined by its normal $n^{\mu }$. 
In order to join two space-time manifolds along $\Sigma$ without a surface 
layer two conditions have to be met \cite{I}: The induced spatial metric 
$h_{ij}\equiv g_{ij} + n_i n_j$ and the extrinsic curvature $K_{ij}$ 
should be continuous on $\Sigma$. The extrinsic curvature is defined as:
\begin{equation}
\label{defK}
K_{ij}=-\frac{1}{2}{\cal L}_n h_{ij},
\end{equation}
where ${\cal L}_n$ denotes the Lie derivative with respect to the 
normal $n^{\mu}$. In order to compute $K_{ij}$ the system of 
coordinates (i.e. the gauge) and the vector $n^{\mu }$ (i.e. the surface 
of transition) have to be specified. Different choices for $n^{\mu}$ lead 
to inequivalent junction conditions. Our derivation of the joining
conditions differs from the derivation in \cite{DM}, where the joining
conditions are calculated in a coordinate system adapted to the surface 
of the transition. For a more general coordinate system the joining conditions
have been obtained by a gauge transformation. 
\par
As a simple example we can apply the previous rules to the background 
model. The surface $\Sigma $ is defined by $q_0(\eta _0)=0$ and the 
components $n_{\mu }$ are given by: $n_0=-a$, $n_i=0$. It is then 
straightforward to show that $K^i{}_j(\eta )=-H(\eta )\delta ^i{}_j$. The 
continuity of the induced spatial metric leads to $\lim_{\epsilon \to 0} 
[a(\eta_\sigma + \epsilon) - a(\eta_\sigma - \epsilon)] \equiv 
[a]_\pm = 0$ whereas the 
continuity of the extrinsic curvature amounts to $[a']_\pm =0$. From the 
Friedmann equations we see that the energy
density can not have a jump, whereas the pressure may jump. Let us turn now 
to the case of scalar perturbations.

\subsection{Scalar perturbations}

The perturbed transition hypersurface is defined by $q_0(\eta_\Sigma) +
\delta q(\eta_\Sigma, x^i) = 0$. From the last expression we immediately 
get that the transition now occurs at time $\eta _{\Sigma }=
\eta _0+\delta \eta =
\eta _0-\delta q/q_0'$. In addition, the normal of $\Sigma $ now reads:
\begin{equation}
\label{ns}
n_0 = - a (1 + \phi), \qquad n_i = - a {\partial_i \delta q \over q_0'}\ , 
\end{equation}
where the scalar perturbations are parameterized by the line 
element (\ref{ds2}). First we must write that the perturbed induced 
metric $h_{ij}$ is continuous on $\Sigma $. Expressing this condition 
for diagonal and off-diagonal terms leads to:
\begin{eqnarray}
\label{psiE1}
[\psi(\eta_\Sigma)]_\pm &=& [\psi + {\cal H} \delta q/q_0']_\pm(\eta )= 0 ,\\
\label{psiE2}
{ }[E(\eta_\Sigma)]_\pm &=& [E(\eta )]_\pm =0.
\end{eqnarray}
Second, we must compute the perturbed extrinsic curvature for the vector 
whose components are given in Eq. (\ref{ns}). We obtain the following result:
\begin{equation}
\label{Kns}
\delta K^i{}_j(\eta )=\frac{1}{a}[(\psi '+{\cal H} \phi) \delta^i{}_j + 
( B - E' + \delta q/q_0')^{|i}{}_{|j}] \ ,
\end{equation}
and therefore, on the surface of transition $\Sigma $, $\delta K^i{}_j$ 
takes the value:
\begin{equation}
\label{KSig}
\delta K^i{}_j(\eta _{\Sigma })=
\frac{1}{a}\{[\psi '+{\cal H} \phi+
({\cal H}'-{\cal H}^2)\frac{\delta q}{q_0'}] \delta^i{}_j + 
(B-E'+\delta q/q_0')^{|i}{}_{|j}\} \ .
\end{equation}
Let us notice that the extrinsic curvature on the hypersurface 
$\Sigma $ is a gauge-invariant quantity. The two other (gauge-invariant) 
junction conditions are easily deduced from the previous expression and read:
\begin{eqnarray}
\label{JCK}
[\psi '+{\cal H}\phi +({\cal H}'-{\cal H}^2)\frac{\delta q}{q_0'}]_\pm(\eta ) 
&=& 0 ,  \\
\label{JCK2}
[B-E'+\frac{\delta q}{q_0'}]_\pm(\eta ) &=& 0.
\end{eqnarray}
These are the same conditions as obtained in Ref.~\cite{DM}. 
We have shown that these are also valid for all ($K = 0,\pm 1$) FLRW models.
\par
We can also establish what are the junction conditions if one chooses 
to match the perturbations on a surface of constant time as it was done in 
Ref.~\cite{G1}. Since $\delta q$ no longer depends on spatial coordinates, 
$\partial _i\delta q=0$, the normal to the surface of transition is 
now given by:
\begin{equation}
\label{neta}
n_0=-a(1+\phi ), \qquad n_i=0,
\end{equation}
and the extrinsic curvature takes on the form:
\begin{equation}
\label{Keta}
\delta K^i{}_j(\eta )=\frac{1}{a}[(\psi '+{\cal H} \phi) \delta^i{}_j + 
(B-E')^{|i}{}_{|j}] \ .
\end{equation}
Requiring the continuity of $h_{ij}$ and $K^i{}_j$ given by (\ref{Keta})
at $\eta_\Sigma = \eta_0$, a surface of constant time, and using
Eq.\ (\ref{gitosyn}) leads to the joining conditions in the synchronous gauge:
\begin{equation}
\label{jsg}
[h]_\pm = [h']_\pm = [h_l]_\pm = [h_l']_\pm = 0 \ .
\end{equation}
These are exactly the conditions that have been used in Ref.~\cite{G1}.
The joining conditions (\ref{psiE1}),(\ref{psiE2}), (\ref{JCK}), 
and (\ref{JCK2}) are not equivalent
to the conditions (\ref{jsg}). This confirms that for a sharp transition 
the choice of the surface of matching is crucial.

However, all joining hypersurfaces are equivalent if $[p]_\pm = 0$. This can be 
easily seen by considering a hypersurface given by (\ref{ns}) in the
longitudinal gauge and then make gauge transformations to any other gauge.
Thus, $B^{\rm (LG)}=E^{\rm (LG)}=0$ and $[\delta q^{\rm (LG)}/q_0']_\pm = 0$
from (\ref{JCK2}). From $[p]_\pm = 0$ and the Friedman equations 
$[{\cal H}']_\pm =0$ follows. The joining conditions (\ref{psiE1}) and 
(\ref{JCK}) imply
\begin{equation} 
[\psi^{\rm (LG)}]_\pm = [\psi^{\rm (LG)\prime} + {\cal H} \phi^{\rm (LG)}]_\pm 
= 0 \ . 
\end{equation}
For a perfect fluid the equation of motion $\phi^{\rm (LG)} = 
\psi^{\rm (LG)}$  reduces the joining conditions to the continuity of the 
metric perturbations and its derivatives. Consider now all hypersurfaces that
are related to the zero shear (longitudinal gauge) hypersurface by the gauge 
transformations (\ref{2-6}) where $\xi^0,\xi \in C^{(2)}$. Then from the 
gauge transformations (\ref{2-6}) and the junction conditions 
it follows that all metric perturbations and its derivatives have 
to be continuous in any gauge that is smoothly connected to the 
longitudinal gauge. Therefore, all these hypersurfaces are equivalent to 
the constant time hypersurface (\ref{neta}) if the pressure does not jump.
We conclude that Grishchuk's joining conditions are correct, provided 
$[p]_\pm =0$. We argue in the next section that this is the case in his 
reheating transition.

\subsection{Vector perturbations}

Let us define the line element for vector perturbations to read 
(here again we use the notations of Ref.~\cite{MFB} except for the 
signature of the metric):
\begin{equation}
\label{ds2v}
{\rm d}s^2 = a(\eta )^2\{-{\rm d}\eta^2 - 2 S_i{\rm d}x^i
{\rm d}\eta + [\gamma_{ij} + F_{i|j} + F_{j|i}]{\rm d}x^i{\rm d}x^j\} \ , 
\end{equation}
with $S_i$ and $F_i$ being transverse vectors, i.e.~$S_i^{\ |i} = 
F_i^{\ |i} = 0$. Fictitious perturbations can be generated by performing 
the infinitesimal change of coordinates:
\begin{equation}
\label{rotgauge}
\bar{\eta }=\eta , \qquad \bar{x}^i=x^i+\zeta ^i,
\end{equation}
where $\zeta ^i{}_{|i}=0$. Under this transformation $S_i$ and $F_i$ change 
according to the equations:
\begin{equation}
\label{SFgauge}
\bar{S}_i=S_i-\zeta _i', \qquad \bar{F}_i=F_i+\zeta _i \ .
\end{equation}
Therefore, we introduce the gauge-invariant dragging 
potential \cite{B1} defined by 
\begin{equation}
\label{Xi}
\Xi_i \equiv S_i+F_i' \ .
\end{equation}
\par
Since there is no possible vector contribution to the normal
of the spatial hypersurface $\Sigma$, the components of the normal are 
simply those of the background, i.e. $n_0=-a$, $n_i=0$. 
{}From the continuity 
of the induced metric we get that $[F_i]_\pm(\eta )=0$ and from the expression 
of the extrinsic curvature
\begin{equation}
\label{Krot}
\delta K^i{}_j=\frac{1}{2a}(\Xi_j{}^{|i}+\Xi^i{}_{|j}),
\end{equation}
we deduce that the second junction condition is: $[\Xi_i]_\pm(\eta )=0$. 
These junction conditions are gauge-invariant. In the synchronous gauge, 
they simply reduce to the continuity of the metric and its derivative.
In the gauge-invariant formulation we have to demand the continuity of
$\Xi_i$ only. The fact that there is only a single condition reflects 
the very different behavior of rotational perturbations compared to density
perturbations or gravitational waves. For perfect fluids the
equation of motion for $\Xi_i$ reads 
\begin{equation}
\Xi'_i + 2{\cal H} \Xi_i = 0 \ ,
\end{equation}  
giving rise to a decaying solution $\propto 1/a^2$ fixed by one initial 
condition only.

\subsection{Tensor perturbations}

Finally, we treat the case of the gravitational waves. The perturbed tensor 
line element is given by (\ref{ds2t}). As for the vector case, the normal 
to the surface of transition is simply the one of the background. The 
continuity of the induced metric leads to $[h_{ij}]_\pm(\eta )=0$ and a 
straightforward calculation of the extrinsic curvature,
\begin{equation}
\label{Kgw}
\delta K^i{}_j=\frac{1}{2a}(h^i{}_j)',
\end{equation}
shows that the derivative of the metric must be continuous as well, 
namely $[h_{ij}']_\pm(\eta )=0$. There are no gauge dependences in 
this sector anyhow.

\section{A smooth transition}

In this section, we analyze the approach taken by Grishchuk 
in Ref. \cite{G1}. He assumed that during inflation, the scale factor 
is given by $a \propto |\eta|^{1+\beta}$ with $1+\beta <0$. During this  
stage, the function $\gamma (\eta )$ is constant and equal to 
$(2+\beta )/(1+\beta )$. Then, instead of matching directly inflation to the 
radiation stage characterized by $a(\eta )\propto (\eta -\eta _e)$, 
$\gamma =2$, he introduced a smooth transition in between. Physically, this 
smooth transition represents the reheating of the Universe. It 
begins at $\eta =\eta _1-\epsilon$ (the end of inflation) and ends at 
$\eta =\eta _1+\epsilon$ (the beginning of radiation). Note that the 
parameter $\epsilon$ introduced above is different from the one used 
in Ref. \cite{G1}. Here, $\epsilon$ is small compared to $\eta_1$ because 
we assume that reheating is fast. In the limit $\epsilon$ goes 
to zero, we recover the sharp transition considered before for which 
$\gamma (\eta )$ becomes an Heaviside function jumping from 
$(2+\beta )/(1+\beta )$ to $2$. In the case of a smooth transition, without 
taking into account all the details of the reheating process, we do not 
know how the scale factor evolves between $\eta _1-\epsilon$ and 
$\eta _1+\epsilon$. It is clear that the function $a(\eta )$ [and 
therefore $\gamma (\eta )$] is probably complicated for this 
stage of the evolution. The idea of Ref. \cite{G1} was to assume 
that the function $\gamma (\eta )$ is given by:
\begin{equation}
\label{interpolgam}
\gamma(\eta) = {4 + 3 \beta\over 2(1 + \beta)} + {\beta\over 2(1 + \beta)} 
\tanh\left( \eta - \eta_1\over s\right) \ ,
\end{equation}
where $s$ is a parameter controlling the sharpness of the transition. This 
equation holds for inflation and reheating, i.e. for $\eta $ between 
$-\infty $ and $\eta _1+\epsilon$. Such a 
postulated behavior for $\gamma (\eta )$ can be justified with the help 
of the following two arguments. First, 
we have $\gamma \simeq (2+\beta )/(1+\beta )$ when $\eta <\eta _1
-\epsilon$ (the parameter $s$ must be chosen such that the $\tanh $ reaches 
quickly the value $-1$; it is sufficient to take $|\epsilon/s|\gg 1$) and we 
recover the 
fact that $\gamma (\eta )$ is constant during inflation. During reheating 
$\gamma (\eta )$ smoothly passes from $(2+\beta )/(1+\beta )$ to its 
exiting value $\gamma (\eta _1+\epsilon)\approx 2$. Second, more physically, 
$\gamma (\eta )$ is related to $w$ by the equation 
$p/\rho=w=-1+(2/3)\gamma (\eta )$. The introduction of the expression 
(\ref{interpolgam}) in the last equation reproduces the expected 
behavior of $w$ in a reasonable model. Therefore, Eq. (\ref{interpolgam}) 
gives a reasonable approximation of the real (exact) complicated function 
$\gamma (\eta )$ even if details of the reheating 
process cannot be taken into account in such a simple approach.
\par
{}From the previous discussion, it is clear that $\gamma (\eta )$ is 
always a continuous function, even at $\eta =\eta _1+\epsilon$ where 
the explicit joining was performed in Ref. \cite{G1}\footnote{ 
In this paper $\eta _1$ was used instead of $\eta _1+\epsilon$ to denote
the end of reheating. It is very important to distinguish these two events.}. 
This means that $[p]_{\pm}$ vanishes, see Eq. (\ref{2-17}). 
In Ref. \cite{DM}, Deruelle and Mukhanov criticized the calculations 
done in Ref. \cite{G1} by means of the smooth transition described before, 
arguing that the junction conditions were not taken into account 
properly. We have shown in the previous section that if the 
pressure is continuous, the two sets of matching conditions, 
Eqs.\ (\ref{psiE1}), (\ref{psiE2}), (\ref{JCK}), (\ref{JCK2}) and 
Eqs.\ (\ref{jsg}) are equivalent. Therefore, the claim of Deruelle and 
Mukhanov is not appropriate. For a smooth transition, the matching 
conditions used by Grishchuk are perfectly justified since they 
coincide with the ones derived in Ref. \cite{DM}. The argument of 
Deruelle and Mukhanov would be correct if the transition were sharp 
and $\gamma (\eta )$ discontinuous at $\eta =\eta _1+\epsilon$. 
Moreover, the exiting values of the 
functions $\gamma (\eta )$ and $\gamma '(\eta )$ at the 
joining reheating-radiation are:
\begin{equation}
\label{exit}
\gamma (\eta _1+\epsilon)=2, \qquad \gamma '(\eta _1+\epsilon)=0,
\end{equation}
in contradiction with the claims of Ref. \cite{DM} but in accordance with 
what is written by Grishchuk.
\par
The next step would be to solve Eq. (\ref{2-40}) for $\gamma (\eta )$ 
given by Eq. (\ref{interpolgam}). It was shown in Ref. \cite{G1} that the 
integration of Eq. (\ref{interpolgam}) can be performed and provides us 
with the function $\alpha (\eta )$. However, obtaining the corresponding 
$a(\eta )$ is not possible. This is not a problem 
since the potential $(a\sqrt{\gamma })''/(a\sqrt{\gamma })$ depends only 
on $\gamma $, $\alpha $ and their derivatives. However, even the simple 
form (\ref{interpolgam}) is to complicated to allow a 
direct integration of Eq. (\ref{2-40}). Nevertheless, we can follow the 
evolution of $\mu $ through inflation and reheating. For 
$\eta <\eta _1-\epsilon$, $\gamma (\eta )$ is a constant and the 
equation (\ref{2-40}) can be solved. The solution reads:
\begin{equation}
\label{bessel}
\mu = (k\eta)^{1/2} [ A_1 J_{\beta + \frac12}(k\eta) + 
A_2 J_{-\beta - \frac12}(k\eta)] \ .
\end{equation}
This solution is the same as for gravitational waves. The initial 
conditions $A_{1,2}$ are fixed by the quantum-mechanical generation of
the density and metric fluctuations: 
see the Appendix. The $A_{1,2}$ differ from the $\tilde{A}_{1,2}$ 
and $\bar{A}_{1,2}$ introduced previously. From Eq.~(\ref{bessel}),
we can determine the value of $\mu (\eta )$  just before reheating:
\begin{equation}
\label{besselexp}
\mu (\eta_1 - \epsilon) \simeq {A_1 \over 2^{\beta+\frac12}
\Gamma(\beta +\frac32) } [k(\eta_1 - \epsilon)]^{\beta + 1} \simeq
{A_1 \over 2^{\beta+\frac12}
\Gamma(\beta +\frac32) } (k\eta_1)^{\beta + 1} \ ,
\end{equation}
because $k\eta_1 \ll 1$ and $\epsilon \ll \eta_1$. Between 
$\eta _1-\epsilon$ and $\eta _1+\epsilon$ the function $\gamma (\eta )$ is 
no longer a constant and the solution (\ref{bessel}) can no longer be 
used. In order to evolve 
$\mu$ through the reheating transition we use the 
superhorizon solution $\mu \sim a\sqrt{\gamma}$ to obtain 
\begin{equation}
\label{mureh}
\mu(\eta_1 + \epsilon) \simeq 
\frac{\mu(\eta_1 - \epsilon)}{a(\eta_1 - \epsilon)
\sqrt{\gamma(\eta_1 - \epsilon)}} a(\eta_1 + \epsilon)  
\sqrt{\gamma(\eta_1 + \epsilon)} \simeq \mu(\eta_1 - \epsilon) 
\sqrt{2\over \gamma_i} \ ,
\end{equation}
because $a(\eta _1+\epsilon)\approx a(\eta _1-\epsilon)$. 
$\gamma _i$ is the value of $\gamma (\eta )$ during inflation. 
This relation should be compared to Eq. (81) and to the relation 
$\mu |_{\eta _1-0}=\mu |_{\eta _1+0}$ below Eq. (48) of Ref. \cite{G1}. 
{}From Eq. (\ref{mureh}), it is clear that the ratio 
$\mu (\eta _1+\epsilon)/\mu(\eta _1-\epsilon)$ is not $1$ but 
proportional to $1/\sqrt{\gamma_i}$. This factor is huge when $\gamma_i$ 
is close to $0$ (de Sitter case). Therefore, the mistake in 
Ref. \cite{G1} was not due to the use of wrong junction conditions but to 
the fact that the function $\mu(\eta )$ was not evolved 
correctly through the reheating transition: actually $\gamma (\eta _1-
\epsilon)\neq \gamma (\eta _1+\epsilon)$ implies 
$\mu(\eta _1-\epsilon)\neq \mu (\eta _1+\epsilon)$. 
\par
The value of $\Phi$ for superhorizon modes during inflation
is obtained by inserting (\ref{bessel}) into (\ref{2-41}), using the 
Taylor expansion of the Bessel functions and substituting 
(\ref{besselexp}) in the corresponding expression. The result reads:
\begin{equation}
\label{poteinf}
\Phi_i \simeq - {\beta + 1\over 2\beta + 3} {\sqrt{\gamma _i}\over 2} 
{\mu(\eta_1 -\epsilon)\over a(\eta_1)} \ .
\end{equation} 
This equation will be used below.
\par
Let us turn now to the second transition, i.e. the transition 
radiation-matter taking place at equality. The matter era is described 
by $a(\eta )\propto (\eta -\eta _m)^2$ and $\gamma =3/2$. In principle, 
one should do the same interpolation at equality as was done at 
reheating. In Ref. \cite{G1}, this was not done and the second 
transition was treated as a sharp 
transition. As was shown in \cite{DM}, the joining
conditions (\ref{jsg}) are not correct in general (i.e., for any residual gauge
fixing). However, if one specifies
the synchronous gauge in the matter dominated epoch to be the comoving one,
then the joining conditions (\ref{jsg}) are fine at the equality transition 
\cite{DM}. That is because the density contrast vanishes at the leading 
order [i.e., it is proportional to $(k\eta)^2$].
This was actually done in \cite{G1}. Neglecting the decaying mode leads to 
the solution
\begin{equation}
\label{solh}
h = C_1\ , \qquad h_l = {1\over 10} C_1 k^2 (\eta - \eta_m)^2 \ .
\end{equation}
The coefficient $C_1$ is related to $\mu(\eta_1 + \epsilon)$, the value
of $\mu$ at the end of reheating, by (see Eq. (84) of Ref. \cite{G1}):
\begin{equation}
\label{C1}
C_1 \simeq {1\over\sqrt{2} a(\eta_1)} \mu(\eta_1 + \epsilon) \ .
\end{equation}
Therefore, everything is known at the matter stage. Let us calculate the 
Bardeen's gauge-invariant potential. Inserting Eq. (\ref{solh}) in the 
formulae (\ref{2-30}) -- (\ref{2-32}) gives:
\begin{equation}
\label{Phim}
\Phi_m \simeq - \frac{3}{10} C_1 = -\frac{3}{10\sqrt{2}a(\eta _1)}
\mu(\eta _1+\epsilon).
\end{equation}
We may now take into account Eqs.~(\ref{poteinf}) and (\ref{Phim}) to arrive 
at
\begin{equation}
\label{ratiomi}
{\Phi_m\over \Phi_i} \simeq \frac{2\beta +3}{1+\beta }
\frac{3}{5\sqrt{2\gamma _i}}
\frac{\mu(\eta _1+\epsilon)}{\mu (\eta _1-\epsilon)}
\simeq \frac25 {2\beta + 3 \over \beta + 1} {1\over 1+w_i} \ .
\end{equation}
Thus, we obtained the `standard result' (\ref{sr}) entirely within
the synchronous gauge, without any reference to the constancy of $\zeta$ or 
the joining conditions of Deruelle and Mukhanov. From the last expression, 
it is clear that it is crucial to evaluate correctly the ratio 
$\mu (\eta _1+\epsilon)/\mu (\eta _1-\epsilon)$.
\par
Finally, it is also interesting to calculate the ratio of $\Phi$ and 
$h_{\rm gw}$ at superhorizon scales today. This quantity is of relevance for 
observations since it is related to the ratio of gravitational 
waves to density perturbations contributing to the CMBR quadrupole
anisotropy. The leading term of the `growing' mode 
of the gravitational waves is constant in time, thus its value today 
is equal to its value at time $\eta =\eta _1-\epsilon $. 
Therefore we obtain:
\begin{equation}
h_{\rm gw}({\rm today}) \simeq {\mu_{\rm gw}(\eta_1 - \epsilon)\over 
a(\eta_1 - \epsilon)}
\approx {A_1^{\rm gw}\over A_1} {\mu(\eta_1 - \epsilon)\over 
a(\eta_1)} \ ,
\end{equation}
where we used the fact that the equations of motion of $\mu$, 
Eq.~(\ref{2-40}), and $\mu_{\rm gw}$, Eq.~(\ref{eomgw}),
are the same during inflation (as long as $\gamma$ is almost constant)
until the onset of reheating at $\eta_1 - \epsilon$. From Eqs.~(\ref{app24}) 
and (\ref{app26}) of the Appendix, it follows that the rms amplitude
of gravitational waves, which takes into account both polarizations, reads: 
\begin{equation}
\label{hrmstoday}
h_{\rm rms} \simeq \frac{\sqrt{2}}{\pi}\left|\frac{A_1^{\rm gw}}{A_1}
\right| \left|\frac{\mu (\eta _1-\epsilon )}{a(\eta _1)}\right| k^{\frac 32}\ .
\end{equation}
On the other hand, Eqs. (\ref{mureh}), (\ref{Phim}) and (\ref{app15}) imply 
that:
\begin{equation}
\label{phirmstoday}
\Phi_{\rm rms} \simeq \frac{3}{20\pi}\sqrt{\frac{2}{\gamma _i}}
\left|\frac{\mu (\eta _1-\epsilon )}{a(\eta _1)}\right| k^{\frac 32}.
\end{equation}
Therefore, the value of the ratio $h_{\rm rms}/\Phi_{\rm rms}$ today is:
\begin{equation}
\label{ratiotoday}
\frac{h_{\rm rms}}{\Phi_{\rm rms}} \simeq \frac{20}{3}\sqrt{\gamma _i}
\left|\frac{A_1^{gw}}{A_1}\right| = \frac{20}{\sqrt{6}}\sqrt{1+w_i},
\end{equation}
where the equations (\ref{app25}) of the Appendix have been used.
This finally proofs that the closer the inflationary epoch is to the de Sitter
space-time, the less important are large-scale 
gravitational waves in the CMBR today.

\section{Conclusion}

In Ref.~\cite{G1} Grishchuk claimed that the magnitude of superhorizon 
scalar metric perturbations is most likely smaller than the amplitude of 
superhorizon gravitational waves. We have shown in Sec.~VI that this result 
is wrong, because the time evolution of the scalar metric perturbation
through the (smooth) reheating transition was not calculated correctly. 
With the appropriate correction we recover the `standard result' for
the rms amplitudes at superhorizon scales
\begin{equation}
\left. {h_{\rm rms}\over \Phi_{\rm rms}} \right|_{\rm today} = 
\frac{20}{\sqrt{6}} \sqrt{1 + w_i}
\sim \left.{m_{\rm Pl}V_{,\varphi}\over V}\right|_{\rm slow-roll}\ ,
\end{equation}
where the slow-roll approximation is valid if $\gamma _i \ll 1$.
However, in the limit $\gamma_i \to 0$ linear perturbation theory 
breaks down since Eq.~(\ref{phirmstoday}) blows up. Thus, for 
power-law inflation, 
the model which we considered in detail, one cannot make the slow-roll 
approximation arbitrarily precise by making $\gamma_i$ arbitrarily small.

Recently, Deruelle and Mukhanov \cite{DM} corrected the result of \cite{G1}
within the framework of sharp transitions. We rederived their joining
conditions in Sec.~V and extended them to non-flat FLRW models ($K\neq 0$). 
Moreover,
we derive the joining conditions for the vector and tensor
perturbations. According to Deruelle and Mukhanov, Grishchuk
made two mistakes: He took the wrong joining conditions and he used the
wrong equation of state (expressed in terms of $\gamma$) at the reheating
transition. However, Grishchuk introduced a tanh to interpolate the pressure 
between the inflation and radiation epochs. Therefore, both his joining 
conditions at the reheating transition and the equation of state after 
reheating have been used correctly. 

A commonly used derivation of the `standard result' has been criticized
by Grishchuk in Refs.~\cite{G2,G3}. This derivation is based on the
conservation of certain quantities for superhorizon modes. These `conservation
laws' are essential in Refs.~\cite{BST,BK,L,B2}. 
We have investigated Grishchuk's arguments in Sec.~III and IV, where we have 
shown that his criticisms are not correct.

{\it Note added:} After our work was finished, a paper by 
M. Goetz (astro-ph/970427) appeared. In this paper he independently reaches 
one of the conclusions of our paper, namely that Grishchuk's claim on 
the emptiness of the `conservation law' is wrong.

\acknowledgements

We would like to thank N. Deruelle, J. C. Fabris, L. P. Grishchuk, 
L. Bel, B. Linet, V. F. Mukhanov, C. Schmid, Ph. Spindel, and P. Widerin 
for valuable discussions. D.S. thanks the Swiss National Science Foundation 
for financial support. J.M. would like to thank the ``Laboratoire 
de Cosmologie et Gravitation Relativistes'' where part of this 
work has been done. 

\appendix

\section*{Quantization}

In this appendix, we briefly review how the perturbations are 
generated quantum-mechanically in the early Universe. This mechanism 
fixes the initial conditions, i.e.~the coefficients 
$A_1^{\rm gw}$, $A_2^{\rm gw}$ and $A_1$, $A_2$ in 
Eqs.~(\ref{solgw}) and (\ref{bessel}). It has been emphasized in the text 
how crucial the precise values of these coefficients are to obtain 
of the final (standard) result.
\par
Let us first consider density perturbations. The normalization of the 
(perturbed) scalar field operator is fixed by the uncertainty principle 
of Quantum Mechanics. In a spatially flat FLRW  model, this leads 
to the following expression:
\begin{equation}
\label{app1}
\delta \hat{\varphi }(\eta ,{\bf x})\equiv \frac{1}{(2\pi )^{3/2}}
\int {\rm d}{\bf k}
\hat{\varphi }_1(\eta ,{\bf k})e^{i{\bf k}\cdot {\bf x}}=
\frac{\sqrt{\hbar }}{a(\eta )}
\frac{1}{(2\pi )^{3/2}}
\int \frac{{\rm d}{\bf k}}{\sqrt{2k}}
[c_{\bf k}(\eta )e^{i{\bf k}\cdot {\bf x}}
+c_{\bf k}^{\dag}(\eta )e^{-i{\bf k}\cdot {\bf x}}],
\end{equation}
where $c_{\bf k}$ and $c_{\bf k}^{\dag }$ are the annihilation 
and creation 
operators satisfying the usual commutation relation. 
This equation agrees with Eq.~(95) of Ref.~\cite{G1}. The initial conditions 
are determined by demanding that at some initial 
time $\eta _0$ (e.g., at the beginning of inflation), the scalar field 
be placed in the vacuum state: $c_{\bf k}(\eta =\eta _0)|0\rangle=0$. 
Then, solving the Heisenberg equation of motion shows that the operator 
$\hat{\varphi }_1(\eta ,{\bf k})$ can be written as:
\begin{equation}
\label{app2}
\hat{\varphi }_1(\eta ,{\bf k})=
\frac{\sqrt{\hbar }}{a(\eta )}
[c_{\bf k}(\eta _0)\frac{u_k+v_k^*}{\sqrt{2k}}
+c_{-{\bf k}}^{\dagger}(\eta _0)\frac{u_k^*+v_k}{\sqrt{2k}}] \ ,
\end{equation}
where the function $(u_k+v_k^*)(\eta )$ satisfies:
\begin{equation}
\label{app3}
(u_k+v_k^*)'' + (k^2 - \frac{a''}{a})(u_k+v_k^*) = 0 \ .
\end{equation}
The initial conditions translate into the statement that $u_k(\eta _0)=1$ and 
$v_k(\eta _0)=0$. In the high frequency regime, this implies the 
asymptotic behavior $\lim _{k\rightarrow +\infty }(u_k+v_k^*)=
e^{-ik(\eta -\eta _0)}$.
\par
The normalization of the perturbed scalar field fixes automatically 
the normalization of the scalar perturbations of the metric since 
they are linked through Einstein's equations. In the high frequency 
limit, this link is expressed through the formula: 
\begin{equation}
\label{app4}
\lim _{k\rightarrow \infty} \hat{\mu } (\eta ,{\bf k})=-\sqrt{2\kappa}
a\hat{\varphi }_1(\eta ,{\bf k})\ ,
\end{equation}
which follows most easily from Eqs.~(\ref{vMmu}) and (\ref{vM}). This 
equation is the same as written at the bottom of p. 7168 
of Ref.~\cite{G1}. This allows to find immediately the asymptotic behavior 
of the operator $\hat{\mu } (\eta,{\bf k})$:
\begin{equation}
\label{app5}
\lim _{k\rightarrow \infty }\hat{\mu }(\eta ,{\bf k})=
-4\sqrt{\pi }l_{\rm Pl}[c_{\bf k}(\eta _0)
\frac{e^{-ik(\eta -\eta _0)}}{\sqrt{2k}}+c_{-{\bf k}}^{\dagger }(\eta _0)
\frac{e^{ik(\eta -\eta _0)}}{\sqrt{2k}}],
\end{equation}
where $l_{\rm Pl}=(G\hbar)^{1/2}$ is the Planck length. For simple 
models where the scale factor is $a(\eta )=
l_0|\eta |^{1+\beta }$, the exact solution for $\hat{\mu } (\eta ,{\bf k})$ 
can be expressed as:
\begin{equation}
\label{app6}
\hat{\mu }(\eta ,{\bf k})=(k\eta )^{1/2}
[\hat{A}_1({\bf k})J_{\beta +1/2}(k\eta )+\hat{A}_2({\bf k})
J_{-(\beta +1/2)}(k\eta )],
\end{equation}
where $J_{\pm (\beta +1/2)}$ is a Bessel function of order 
$\pm (\beta +1/2)$. The two last equations imply that the (so far 
arbitrary) operators $\hat{A}_1({\bf k})$ and $\hat{A}_2({\bf k})$ 
are given by:
\begin{eqnarray}
\label{app7}
\hat{A}_1({\bf k}) &=& \frac{i \sqrt{8}\pi l_{\rm Pl}}{\cos \beta \pi}
\frac{1}{\sqrt{2k}}
[e^{i(k\eta _0+\frac{\pi \beta }{2})}c_{{\bf k}}(\eta _0)
-e^{-i(k\eta _0+\frac{\pi \beta }{2})}c_{-{\bf k}}^{\dagger}(\eta _0)],
\\
\label{app8}
\hat{A}_2({\bf k}) &=&
-\frac{\sqrt{8}\pi l_{\rm Pl}}{\cos \beta \pi} \frac{1}{\sqrt{2k}}
[e^{i(k\eta _0-\frac{\pi \beta }{2})}c_{\bf k}(\eta _0)
+e^{-i(k\eta _0-\frac{\pi \beta }{2})}c_{-{\bf k}}^{\dagger}(\eta _0)].
\end{eqnarray}
Essentially, these relations agree with Eqs.~(102) of Ref.~\cite{G1}. 
\par
The power spectrum for the Bardeen potential can now be computed. Its 
definition is given in terms of the two-point correlation function for 
$\hat {\Phi }(\eta ,{\bf x})$:
\begin{equation}
\label{app9}
\langle 0|\hat{\Phi }(\eta,{\bf x})\hat{\Phi}(\eta,{\bf x}+{\bf r})|0\rangle 
=\int _0^{\infty }
\frac{{\rm d}k}{k}\frac{\sin kr }{kr}k^3P_{\Phi }(k).
\end{equation} 
Using Eq.~(3.27) which expresses the link between $\mu $ and $\Phi $, we 
find the following expression valid for the simple models evoked 
previously and for long wavelengths:
\begin{equation}
\label{app10}
P_{\Phi}(k)=\frac{l_{\rm Pl}^2}{l_0^2}\frac{\gamma (1+\beta )^2}{2^{2\beta +4}
\cos ^2(\beta \pi )\Gamma ^2(\beta +5/2)}k^{2\beta +1}.
\end{equation}
So far, this is the result during inflation on superhorizon scales.
Although no explicit time dependence is visible in (\ref{app10}),
the value of $P_{\Phi}(k)$ changes during reheating and during the
equality transition. In particular, the spectrum today is equal to: 
$P_{\Phi }({\rm today},k)=T(k)P_{\Phi }({\rm initial},k)$. For super 
horizon modes, the transfer function is given by: 
$\lim _{k \rightarrow 0}T(k)=[3(2\beta +3)]^2/[5(1+\beta )\gamma]^2$.
If space-time during inflation was close to a de Sitter phase, 
then the spectrum today is the well-known 
Harrison-Zeldovich (scale-invariant) spectrum (i.e. $\beta \lesssim -2$).
\par
This result permits us to carry the quantum-mechanical initial 
conditions to the classical level. Let us define the `classical 
spectrum' for the classical quantity $\Phi (\eta ,{\bf x})$ by the 
following expression:
\begin{eqnarray}
\label{app11}
\langle \Phi (\eta ,{\bf x})\Phi (\eta ,{\bf x}+{\bf r})\rangle
&\equiv & \frac{\int _V{\rm d}{\bf x}\Phi (\eta ,{\bf x})\Phi (\eta ,
{\bf x}+{\bf r})}{\int _V{\rm d}{\bf x}} \\
\label{app12}
&=& \int _0^{\infty }\frac{{\rm d}k}{k}\frac{\sin kr}{kr}k^3
P_{\Phi }^{\rm cl}(k),
\end{eqnarray}
where $P_{\Phi}^{\rm cl}(k)$ is given by:
\begin{equation}
\label{app13}
P_{\Phi }^{\rm cl}(k)=\frac{1}{2\pi ^2}|\Phi (\eta ,{\bf k})|^2.
\end{equation}
In this expression, $\Phi (\eta ,{\bf k})$ can be calculated using 
Eqs.~(3.27) and (6.3) where the unknown coefficients $A_1$, 
$A_2$ appear. Requiring that $P_{\Phi }(k)=P_{\Phi }^{\rm cl}(k)$ 
fixes $|A_1|$. This last equation relies on the ergodic 
assumption, namely that ensemble averages are equal to spatial 
averages. In addition, if we have the following behavior 
for the classical $\mu (\eta ,{\bf k})$: 
$\lim _{k\rightarrow +\infty }\mu =-4\sqrt{\pi }l_{\rm Pl}
e^{-ik(\eta -\eta _0)}/\sqrt{2k}$ as it is suggested by Eq. (\ref{app5}), 
then the classical initial conditions $A_1$, $A_2$ are 
completely determined. They read:
\begin{equation}
\label{app14}
A_1 =\frac{i \sqrt{8}\pi l_{\rm Pl}}{\cos \beta \pi}
\frac{e^{i(k\eta _0+\frac{\pi \beta }{2})}}{\sqrt{2k}}, \qquad
A_2 =i A_1 e^{-i\pi \beta }.
\end{equation}
Finally, we define the root mean square value $\Phi_{\rm rms}$ by:
\begin{equation}
\label{app15}
\Phi _{\rm rms}\equiv \sqrt{k^3 P_{\Phi}(k)}.
\end{equation}
This quantity is used at the end of Section VI. 
\par
Let us now consider gravitational waves. The problem is very similar to 
the previous one since the quantization of gravitational waves is 
equivalent to the quantization of two scalar fields (representing the two 
independent degrees of freedom of the wave). Assuming again that the 
initial state is the vacuum leads to the following equation for the 
gravitational wave operator: 
\begin{eqnarray}
\label{app16}
\hat{h}_{ij}(\eta ,{\bf x}) &=& \frac{1}{a(\eta )}\frac{1}{(2\pi )^{3/2}}
\sum _s\int {\rm d}{\bf k}p_{ij}^s({\bf k})\hat{\mu }_{\rm gw}^s(\eta ,{\bf k})
e^{i{\bf k}\cdot {\bf x}} \\
\label{app17}
&=& \frac{4\sqrt{\pi }l_{\rm Pl}}{a(\eta )}
\frac{1}{(2\pi )^{3/2}}
\sum _s\int \frac{{\rm d}{\bf k}}{\sqrt{2k}}p_{ij}^s({\bf k})
[(u_k^s+v_k^{s*})c_{\bf k}^s(\eta _0)e^{i{\bf k}\cdot {\bf x}}
\nonumber \\
&+& (u_k^{s*}+v_k^s)c_{\bf k}^{s\dag}(\eta _0)e^{-i{\bf k}\cdot{\bf x}}].
\end{eqnarray}
In these formulas, $p_{ij}^s({\bf k})$ is the (transverse-traceless) 
polarization tensor and the summation over $s$ represents the summation 
over the two states of polarization of the wave. The polarization tensor
is normalized as $p_{ij}^s({\bf k}) p^{s'\ ij}({\bf k})=2\delta^{ss'}$.
The function $u_k^s +v_k^{s*}$ satisfies Eq. (2.16) (and in 
fact does not depend on the state of polarization $s$). The initial 
conditions are: $u_k^s(\eta _0)=1$ and $v_k^s(\eta _0)=0$. This implies 
the following asymptotic behavior for the operator 
$\hat {\mu }_{\rm gw}^s(\eta ,{\bf k})$:
\begin{equation}
\label{app18}
\lim _{k\rightarrow \infty }\hat{\mu }_{\rm gw}^s(\eta ,{\bf k})=
4\sqrt{\pi }l_{\rm Pl}[c_{\bf k}^s(\eta _0)
\frac{e^{-ik(\eta -\eta _0)}}{\sqrt{2k}}+c_{-{\bf k}}^{\dagger s}(\eta _0)
\frac{e^{ik(\eta -\eta _0)}}{\sqrt{2k}}].
\end{equation}
This equation is similar to Eq. (\ref{app5}). Since the exact solution for 
$\hat{\mu }_{gw}^s(\eta ,{\bf k})$ can be expressed as:
\begin{equation}
\label{app19}
\hat{\mu }_{\rm gw}^s(\eta ,{\bf k})=(k\eta )^{1/2}
[\hat{A}_1^{\rm gw}({\bf k},s)J_{\beta +1/2}(k\eta )+
\hat{A}_2^{\rm gw}({\bf k},s) J_{-(\beta +1/2)}(k\eta )]\ ,
\end{equation}  
this fixes the operators $\hat{A}_1^{\rm gw}({\bf k},s)$ and 
$\hat{A}_2^{\rm gw}({\bf k},s)$:
\begin{eqnarray}
\label{app20}
\hat{A}_1^{\rm gw}({\bf k},s) &=& -\frac{i \sqrt{8}\pi 
l_{\rm Pl}}{\cos \beta \pi}
\frac{1}{\sqrt{2k}}
[e^{i(k\eta _0+\frac{\pi \beta }{2})}c_{\bf k}^s(\eta _0)
-e^{-i(k\eta _0+\frac{\pi \beta }{2})}c_{-{\bf k}}^{\dagger s}(\eta _0)],
\\
\label{app21}
\hat{A}_2^{\rm gw}({\bf k},s) &=& 
\frac{\sqrt{8}\pi l_{\rm Pl}}{\cos \beta \pi} \frac{1}{\sqrt{2k}}
[e^{i(k\eta _0-\frac{\pi \beta }{2})}c_{\bf k}^s(\eta _0)
+e^{-i(k\eta _0-\frac{\pi \beta }{2})}c_{-{\bf k}}^{\dagger s}(\eta _0)] \ .
\end{eqnarray}
The result for gravitational waves is very similar to the result for 
density perturbations.
\par
We can also compute the power spectrum for gravitational 
waves (it involves the calculation of $\langle 0|\hat{h}_{ij}(\eta ,{\bf x})
\hat{h}^{ij}(\eta ,{\bf x}+{\bf r})|0\rangle $). For long wavelengths, we 
obtain:
\begin{equation}
\label{app22}
P_{h}(k) =\frac{l_{\rm Pl}^2}{l_0^2}\frac{1}{2^{2\beta -2}\cos ^2(\beta \pi )
\Gamma ^2(\beta +3/2)}k^{2\beta +1}.
\end{equation}
Interestingly enough, the $k$-dependence of the density perturbations and 
gravitational waves power spectrum is the same. This behavior is a special 
feature of power-law inflation. As for the density perturbations case, 
the calculation of the classical spectrum leads to the determination 
of the classical initial conditions. The classical spectrum is given 
by:
\begin{equation}
\label{app23}
\langle h_{ij}(\eta ,{\bf x})
h^{ij}(\eta ,{\bf x}+{\bf r})\rangle =\int _0^{\infty }
\frac{{\rm d}k}{k}\frac{\sin kr}{kr}k^3P_h^{\rm cl}(k),
\end{equation}
where $P_h^{\rm cl}(k)$ can be expressed as:
\begin{equation}
\label{app24}
P_h^{\rm cl}(k)=\frac{2}{\pi^2}\left|\frac{\mu_{\rm gw}^s(\eta,{\bf k})}{a}
\right|^2.
\end{equation}
Requiring $P_h(k)=P_h^{\rm cl}(k)$ and demanding the same high frequency 
behavior for the quantum and the classical solution 
leads to the determination of 
$A_1^{\rm gw}$, $A_2^{\rm gw}$ which appear as free numbers in the 
expression of $\mu _{\rm gw}(\eta, {\bf k})$, Eq. (\ref{solgw}). We obtain 
the following result:
\begin{equation}
\label{app25}
|A_1^{\rm gw}|=|A_1|\ , \qquad |A_2^{\rm gw}|=|A_2| \ .
\end{equation}
Thus we have derived the result used in Sec.~VI:
$|A_1^{\rm gw}/A_1|=1$. 
\par
Finally, the classical quantity $h_{\rm rms}$ is defined by:
\begin{equation}
\label{app26}
h_{\rm rms}=\sqrt{k^3 P_{h}(k)}.
\end{equation}

\begin{table}
\begin{tabular}{ll}
${\cal H} \equiv a'/a$ & ${\cal H} = a H$, where $H$ is the expansion rate.\\
$\phi,B,\psi,E$ &
Perturbed scalar metric coefficients, see (\ref{ds2}). \\
$\Phi$, $\Psi$ & 
Gauge-invariant scalar metric-perturbation, see (\ref{gimp}). \\
$\sigma$ & 
Proportional to $\Phi$, see (\ref{sigma}). Denoted $u$ in \cite{MFB}. \\
$v_{\rm M}$ &  
Gauge-invariant velocity potential, see (\ref{vM}). 
Denoted $v$ in \cite{MFB}.\\
$S_i, F_i$ & 
Perturbed vector metric coefficients, see (\ref{ds2v}). \\
$\Xi_i$ & 
Gauge-invariant vector metric-perturbation, see (\ref{Xi}). \\
$h_{ij}$ & Perturbed tensor metric coefficient, see (\ref{ds2t}).  
\end{tabular}
\caption{Notation of gauge-invariant formalism.}
\end{table}

\begin{table}
\begin{tabular}{ll}
$\alpha \equiv a'/a$ & $\alpha = a H$, where $H$ is the expansion rate.\\
$\gamma \equiv 1 - \alpha'/\alpha^2$ & 
$\gamma$ is related to $w=p_0/\rho_0$ by $\gamma = [3(1+w)/2](1-K/\alpha^2)$. \\
$h,h_l$ &
Perturbed scalar metric coefficients, see (\ref{ds2Gri}). \\
$u,v$ &
Residual-gauge-invariant scalar metric-perturbation, see (\ref{2-30}). \\
$\mu$ & Proportional to $u$, see (\ref{mu}). 
\end{tabular}
\caption{Notation of synchronous gauge.}
\end{table}

\end{document}